\documentclass[prx,twocolumn,aps,epsf,showpacs,superscriptaddress]{revtex4-1}
\usepackage[pdftex]{graphicx}

\usepackage{dcolumn}
\usepackage{bm}
\usepackage{epsfig}
\usepackage{latexsym} 
\usepackage{amsmath}
\usepackage{amssymb}
\usepackage{color}
\usepackage{array}
\usepackage{bbm}
\usepackage{hyperref}
\usepackage{color}
\usepackage{array}
\usepackage{cancel}
\usepackage{ulem}
\usepackage{booktabs}
\newcolumntype{C}[1]{>{\centering\arraybackslash}m{#1}}
\AtBeginDocument{
\heavyrulewidth=.08em
\lightrulewidth=.05em
\cmidrulewidth=.03em
\belowrulesep=.65ex
\belowbottomsep=0pt
\aboverulesep=.4ex
\abovetopsep=0pt
\cmidrulesep=\doublerulesep
\cmidrulekern=.5em
\defaultaddspace=.5em
}
\usepackage{booktabs}
\newcolumntype{R}[1]{>{\raggedleft\arraybackslash}p{#1}}
\AtBeginDocument{
\heavyrulewidth=.08em
\lightrulewidth=.05em
\cmidrulewidth=.03em
\belowrulesep=.65ex
\belowbottomsep=0pt
\aboverulesep=.4ex
\abovetopsep=0pt
\cmidrulesep=\doublerulesep
\cmidrulekern=.5em
\defaultaddspace=.5em
}

\newcommand{\<}{\langle}
\newcommand{\e}{\varepsilon}
\newcommand{\up}{\uparrow}
\newcommand{\down}{\downarrow}
\renewcommand{\>}{\rangle}
\renewcommand{\(}{\left(}
\renewcommand{\)}{\right)}

\newcommand{\Z}{\mathbb{Z}}
\newcommand{\T}{\mathcal{T}}

\newcommand{\blue}[1]{\textcolor{blue}{#1}}
\newcommand{\header}[1]{\vspace{4pt}\noindent{\bf #1 -- }}

\begin{document}
%\title{An infinite family of dynamical $3d$ topological insulators from time-periodic driving}
\title{An infinite family of $3d$ Floquet topological paramagnets}

\author{Andrew C. Potter}
\affiliation{Department of Physics, University of Texas at Austin, Austin, TX 78712, USA}

\author{Ashvin Vishwanath}
\affiliation{Department of Physics, Harvard University, Cambridge MA 02138, USA}

\author{Lukasz Fidkowski}
\affiliation{Department of Physics and Astronomy, Stony Brook University, Stony Brook, NY 11794, USA}
\affiliation{Kavli Institute for Theoretical Physics, University of California, Santa Barbara, CA 93106, USA}

\begin{abstract}
We uncover an infinite family of time-reversal symmetric $3d$ interacting topological insulators of bosons or spins, in time-periodically driven systems, which we term Floquet topological paramagnets (FTPMs). These FTPM phases exhibit intrinsically dynamical properties that could not occur in thermal equilibrium, and are governed by an infinite set of $\Z_2$-valued topological invariants, one for each prime number. The topological invariants are physically characterized by surface magnetic domain walls that act as unidirectional quantum channels, transferring quantized packets of information during each driving period. We construct exactly solvable models realizing each of these phases, and discuss the anomalous dynamics of their topologically protected surface states.  % including possible surface Floquet enriched topological orders. 
Unlike previous encountered examples of Floquet SPT phases, these $3d$ FTPMs are not captured by group cohomology methods, and cannot be obtained from equilibrium classifications simply by treating the discrete time-translation as an ordinary symmetry. 
%We further explore closely related $3d$ Floquet topological insulators of spinless fermions that are intrinsically interacting, that are neither bosonic nor contained within a band-structure classification.
The simplest such FTPM phase can feature anomalous $\Z_2$ (toric code) surface topological order, in which the gauge electric and magnetic excitations are exchanged in each Floquet period, which cannot occur in a pure 2d system without breaking time reversal symmetry.
\end{abstract}
\maketitle

Low temperature $2d$ systems with a gap exhibit discretely quantized electrical and thermal Hall conductance associated with chiral edge channels that reflect the underlying bulk topology of the many-body state. In the absence of excitations with fractional charge and statistics, there are fundamental minimum values for these topological quantities. For example, the minimal thermal Hall conductance per unit temperature is $\kappa_{0,f}/T = 1$ for (non-superconducting) fermion systems, and $\kappa_{0,b}/T = 8$  for bosonic systems
\footnote{Here, we work in units of $\frac{e^2}{h} L$ where $L = \frac{\pi^2k_B^2}{3e^2}$ is the Lorenz number and $T$ is the temperature}. Strikingly $3d$ time-reversal symmetry (TRS) symmetry protected topological phases (SPTs), such as electronic topological insulators\cite{kane2005z2} (eTIs) and their interacting bosonic counterparts, topological paramagnets\cite{vishwanath2013physics,wang2013boson,wang2014classification,burnell2014} (TPMs), can violate these fundamental constraints, exhibiting anomalous surface states with effectively {\it half} of this minimum Hall quantization. For example, in eTIs this half-integer surface Hall conductance is a direct consequence of the unpaired surface Dirac fermion that is topologically protected by TRS and charge-conservation. This anomalous Hall conductance has been observed by examining domains between opposite surface magnetizations, whose interface carries a single chiral electron mode with electric and thermal Hall conductance twice that of the neighboring magnetic domains~\cite{chang2013experimental}. Equivalently, this surface property corresponds to bulk electromagnetic and gravitational axion angles $\theta_{e,g} = \pi$~\cite{witten1979dyons}, leading to a topological magneto-electric effect~\cite{qi2009inducing}, and exposing unexpected connections between electronic materials and non-perturbative anomalies and dualities of gauge theories~\cite{witten1979dyons,son2015composite,wang2015dual,metlitski2016particle,seiberg2016duality,karch2016particle}. 

These phenomena are, by now, rather well understood in thermal equilibrium settings, even in the presence of strong interactions, due to recent advances in understanding and systematically classifying SPT phases~\cite{kane2005z2,fu2007topological,moore2007topological,kitaev2009periodic,ryo10topological,fidkowski2011topological,turner2011topological,chen2011complete,chen2012symmetry,chen2013symmetry,lu2012theory,vishwanath2013physics,wang2013boson,wang2014classification,wang2014interacting,metlitski2014interaction}. In this paper, we venture beyond this familiar equilibrium setting to investigate new non-equilibrium SPT phases arising in ``Floquet" systems that are subjected to time-periodic driving\cite{oka2009photovoltaic,lindner2011floquet,kitagawa2010topological,jiang2011majorana,rudner2013anomalous,khemani2015phase,von2016phaseI,else2016classification,potter2016classification,roy2016abelian,po2016chiral,roy2016periodic,harper2016stability,po2017radical}. 

In this context, quantum Hall systems with non-zero Chern number are unstable to drive-induced heating~\cite{nandkishore2014marginal,potter2015protection}. Instead, time-periodic driving enables a new set of topological phases with chiral edge dynamics but zero Hall conductance~\cite{rudner2013anomalous}, dubbed chiral Floquet (CF) phases~\cite{po2016chiral}, which can be many-body localized (MBL)~\footnote{Whether stable MBL can occur in dimension higher than one remains remains an important, unsettled matter of principle\cite{de2016stability}. For strong disorder the dynamics will behave as in an MBL system at worst up to super-exponentially long timescale, and possibly forever. Further, quasi-periodic MBL could avoid the potentially problematic rare regions effects~\cite{khemani2017two}. }. 
In the absence of fractional excitations~\cite{gross2012index,po2016chiral,po2017radical}, CF edge channels periodically pump a quantized integer number of quantum states, $p_R$ to the right, and an integer number of states, $p_L$, to the left. This pumping is characterized by a topological invariant $\nu = \log \frac{p_R}{p_L}$, that is the logarithm of a rational fraction, inspiring the name ``rational CF" phases. 
%While superficially similar to integer quantum hall states, CF phases are distinct dynamical phenomena that can only occur far from equilibrium.
%: i) rather than a continuous flow of charge and heat at their boundary, CF phases exhibit a discrete unidirectional pumping of a quantized amount of quantum information during each driving period, and ii) upon combining two of these phases, the rational fractions multiply, in contrast to the additive structure of Hall conductance.  

In this paper, we investigate whether there are $3d$ TRS SPT phases whose surface states can exhibit a ``fraction" of the minimal dynamical chiral invariant, $\nu$, analogous to anomalous Hall conductance and anomalous $\theta$-angle of equilibrium TIs and TPMs.  To avoid technical complications associated with fermion systems, we focus on $3d$ Floquet systems of interacting bosons or spins subjected to a TRS drive. In this context, we uncover an infinite family of $3d$ FSPT phases, which we refer to as Floquet topological paramagnets (FTPM). In analogy to how the surface state of an equilibrium topological paramagnet is effectively a TRS ``half" of the minimal $2d$ integer thermal quantum Hall insulator, the surface states of these $3d$ FTPM phase are effectively ``square-roots" of minimal chiral Floquet (CF) phases~\cite{rudner2013anomalous,po2016chiral,po2017radical}. Just as a magnetic domain wall at the surface of an equilibrium eTI behaves as the edge of an integer quantum Hall phase with odd-integer Hall conductance, a TRS-breaking domain on the surface of the $3d$ DTI phase exhibits the same dynamics as the edge of a $2d$ rational CF phase whose rational topological invariant is not a perfect square (the multiplicative analog of ``odd"). These FTPM phases are governed by an infinite set of dynamical $\Z_2$-valued topological invariants, one for each prime number, or equivalently the positive rationals modulo perfect squares, $\mathbb{Q}_+/\mathbb{Q}_+^2$~\footnote{One might worry about the stability of such an infinite classification, however, we remark that any reasonable lattice model will contain a bounded number, $D$, of states per site, in which case we can only realize phases with invariant up to the largest prime factor of $D$.}. After constructing these invariants, we build solvable lattice models for driven systems that realize each of these $3d$ FTPM phases, and explore their anomalous, topological surface state dynamics.

A complementary perspective on the $3d$ FTPMs is provided by their possible anomalous $2d$ surface topological orders. The simplest example is the $3d$ FTPM phase realized in a spin-1/2 lattice model, whose surface has chiral unitary index $\log \sqrt{2}$. We will show that this phase can exhibit a Floquet enriched $\Z_2$-topological order (Toric code) with emergent gauge electric $e$, magnetic, $m$ that get periodically interchanged, $e\leftrightarrow m$. In a purely $2d$ system, we have previously shown\cite{po2017radical,fidkowski2017interacting} that this $e\leftrightarrow m$ exchange is necessarily accompanied by TRS-breaking radical CF edge state with chiral index $\nu = \pm \frac12\log{2}$. However, the special (anomalous) feature of the $3d$ Floquet phase, is that it enables this surface $e\leftrightarrow m$ exchanging surface topological order to occur in a time-reversal symmetric fashion.

\header{Topological invariants for 3D FTPMs}
We begin by constructing a new dynamical topological invariant for $3d$ FTPMs.
Our setting will be a $3d$ system of interacting bosons (e.g. spins), subjected to a time-dependent Hamiltonian $H(t)$ with period, $T$, $H(t) = H(t+T)$, and associated with the Floquet operator (time-evolution operator for one period):
\begin{align}
U(T) = \hat{T}e^{-i\int_0^TH(t) dt}
\end{align}
where $\hat{T}$ denotes time ordering. In this driven setting, heating can be either avoided by introducing strong disorder to drive the system into a many-body localized (MBL) regime~\cite{nandkishore2015many,lazarides2015fate,abanin2016theory},
or postponed for an exponentially long time by rapid driving~\cite{abanin2015exponentially,abanin2015effective,kuwahara2016floquet,else2016pre}. We will restrict our attention to MBL settings, though we expect our results to extend straightforwardly to pre-thermal systems.  Our focus will be on drives with time-reversal symmetry implemented by an anti-unitary operator $\T$ acting as: $\T H(t)\T^{-1} = H(-t)$ (i.e. $\T U(T)\T^{-1} = U^\dagger(T)$), which acts like $\T^2=1$ on all particles. The latter requirement avoids local Kramers degeneracies that would spoil MBL~\cite{potter2016symmetry}. %After exploring the simpler case of bosons, we will investigate the more subtle case of $3d$ fermion FSPTs.

In the absence of a boundary, the Floquet evolution is MBL, and decomposes into the product of quasi-local unitary operations: $U(T) = \prod_\alpha U_\alpha$, where $U_\alpha$ commute for different $\alpha$, and are exponentially well localized near position $r_\alpha$. There are two distinct ways to define the action of $U(T)$ in the presence of a spatial boundary:
\begin{enumerate} 
\item We can simply truncate the terms in the Hamiltonian $H(t)$ that cross the boundary. 
\item Alternatively,  we could omit the factors of $U_\alpha$ whose position, $r_\alpha$, lies within a finite width region near the boundary.
\end{enumerate}
 Denoting these two bulk-truncated Floquet evolutions as $U_{B_{1,2}}$ respectively, we can then identify the action of $U(T)$ at the boundary by their difference:
\begin{align} 
Y = U_{B_2}^{-1}U_{B_1}
\end{align}
which is exponentially well-localized to the boundary. By definition, the phase is trivially localizable in the absence of symmetries. This ensures that we can write $Y$ as a local Hamiltonian evolution of a 2d Hamiltonian: $Y = Te^{-i\int_0^TH_\text{S}(t)dt}$, where $H_\text{S}(t)$ is exponentially well localized to the surface. Since $U_{B_{1.2}}$ are manifestly TRS, and commute \footnote{More precisely, the non-commutation of these terms is exponentially small in the ratio thickness of $B_2$ divided by the localization length, which can be made arbitrarily small.}, $Y$ is also TRS.  Crucially, however, it might be the case that its generating Hamiltonian $H_\text{S}(t)$ necessarily breaks TRS.  To quantify this obstruction to the existence of a TRS generating Hamiltonian, we first pick any generating Hamiltonian $H_\text{S}(t)$ and from it construct a modified 2D Hamiltonian with enlarged period $2T$:
\begin{align}
H'_\text{S}(t) = \begin{cases} H_S(t) & \text{for } 0<t<T \\ - \T H_S(t-T)\T^{-1} & \text{for } T<t<2T \end{cases}
\label{eq:Hprime}
\end{align}
where the minus sign in the second line makes $H'_\text{S}(t)$ a time-reversal anti-symmetrized version of $H_S$.  Since, $Y$ is TRS on a closed surface. this anti-symmetrization implies $U'_\text{S}(2T) = e^{-i\int_0^{2T} H'_\text{s} dt}= \T Y\T^{-1} Y= 1$, i.e. $U'_\text{S}(t)$, $0 \leq t < 2T$ forms a closed loop in the space of finite depth unitaries. Moreover, since $H'_\text{S}$ is a purely 2D Hamiltonian, we can truncate it to a finite $2d$ disk and compute its chiral unitary edge index, $\nu$, which is equal to the $\nu$ of $H_S$ minus the $\nu$ of its TR-conjugate (due to the TRS anti-symmetrization in Eq.~\ref{eq:Hprime}).
%Since, $Y$ is TRS, on a closed surface, $U'_\text{S}(2T) = e^{-i\int_0^{2T} H'_\text{s} dt}= \T Y\T^{-1} Y= 1$ (i.e. $U'_\text{S}(t)$, $0 \leq t < 2T$ forms a closed loop in the space of finite depth unitaries), 
Since $Y=1$ away from the edge of the disk $\nu$ takes a (log) rational value~\cite{gross2012index,po2016chiral,cirac2017matrix,sahinoglu2017matrix}: 
\begin{align}
\nu(Y) = \log r(Y)
\end{align} 
for some rational $r$. It will be convenient to represent this rational number via its unique prime factorization: 
\begin{align}
r(Y) = \prod_i p_i ^{n_i(Y)},
\end{align}
where $p_i$ is the i$^\text{th}$ prime number, and $n_i\in \Z$. 

Without affecting the bulk dynamics, we can alter $Y$ by attaching a purely $2d$ rational CF phase with chiral unitary index $\nu_{2d} = \sum_i m_i\log p_i$ to the surface. Such surface deformations preserve the unitary loop property of $U'_\text{S}$. Due to the anti-symmetriziation in Eq.~\ref{eq:Hprime}, this changes $\nu[U'_\text{S}]$ by twice that of the attached $2d$ phase, i.e. $r\rightarrow \prod_i p_i^{n_i+2m_i}$. 
Given the completeness of the bosonic CF classification\cite{po2016chiral}, this $2d$ CF alteration is the only way to modify the surface $\nu$ with a finite depth unitary transformation of the surface evolution, such that the integers $n_i$ are well-defined modulo $2$. Each of these integers gives a distinct $\Z_2$ valued topological invariants, $\{n_i(Y)~\text{mod}~2\}$. The infinite set of $\Z_2$ invariants  (one for prime number), can be conveniently expressed via an integer:
\begin{align}
\eta(Y) = \prod_i p_i^{n_i(Y)~\text{mod}~2}.
\end{align}
In this notation, $\eta(Y)$ combines multiplicatively upon composing different TRS unitary evolutions, and should be viewed as an element of the rationals modulo the rationals squared, $\eta \in \mathbb{Q}_+/\mathbb{Q}_+^2$.

\header{Models}
Having uncovered a $\Z_2^\infty$ (equivalently $\mathbb{Q}_+/\mathbb{Q}_+^2$) valued topological invariant for 3D FTPMs, we next show that all values of this invariant are realizable in a local 3D system by constructing explicitly solvable lattice models. We emphasize that these models are {\it not} intended as a realistic proposal for the implementation of these FTPM phases. Rather, they provide a formal proof of existence for FTPM phases with arbitrary invariant $\eta$, and serve as a controlled theoretical platform for investigating their anomalous surface dynamics.

The key physical property of the $3d$ FTPM phases is already visible from the structure of $\eta(Y)$: the interface between one region of the surface governed by $H_\text{S}$ and another governed by $\T H_\text{S}\T^{-1}$ will behave like a CF edge with chiral unitary index $\nu = \log n_i(Y)$. This motivates a ``decorated domain wall" construction\cite{chen2014ddw}, in which magnetic domain walls (DWs) are ``decorated" with chiral Floquet phases having $\nu = \log p$.

Specifically, we consider a $3d$ lattice of spins-1/2, $\vec{\sigma}_i$ that transform under TRS as $\T\vec{\sigma}_i\T^{-1} = \sigma^x_i\vec{\sigma}_i^*\sigma_i^x$. The model also contains $p$-state boson degrees of freedom on a dual lattice with one spin at the center of each boson unit cell, transforming trivially under TRS (see Appendix~\ref{app:model} for details). 
 For a fixed spin configuration, we can identify domains of $\sigma^z=\up$ or $\down$, and define an orientation for the 2D DW surfaces via a normal vector on each cubic face of the ``particle" cubes that points from $\sigma^z = \down$ to $\sigma^z=\up$. Then, our strategy will be to evolve the bosons on each DW with the unitary evolution of a 2D chiral Floquet (CF) phase, with chirality chosen in a right-handed sense with respect the DW orientation.  One can attempt to implement these DDW dynamics by a unitary time evolution of the form:
\begin{align} 
U_\text{DDW} = \hat{T}e^{-i \int_0^T dt \sum_{a,s} \Pi_{a\in \text{DW}_s} H_{\text{CF},s,a}(t)\Pi_{a\in \text{DW}_s}}
\label{eq:uddw}
\end{align}
where, $\Pi_{a\in \text{DW}_s}$ is a projection operator onto spin configurations in which plaquette $a$ resides on a DW with orientation $s=\pm 1$, and $H_{\text{CF},s,a}(t)$ is the (time-dependent) Hamiltonian for a chiral phase of bosons residing on plaquette $a$, with chirality $s=\pm 1$. An explicit lattice-scale implementation of $H_{\text{CF},s,a}$ for an arbitrary DW geometry is given in Appendix~\ref{app:model}. As written, the schematic form Eq.~\ref{eq:uddw} is not manifestly TRS invariant. However, in Appendix~\ref{app:model}, we show that, by breaking the CF evolution on the spin domain into pieces first evolving plaquettes oriented in the x- and y- directions, and subsequently those in the z-direction, we can implement an equivalent unitary evolution in a manifestly TRS sequence of steps.

\begin{figure}[t]
\begin{center}
{\includegraphics[width=0.35 \textwidth]{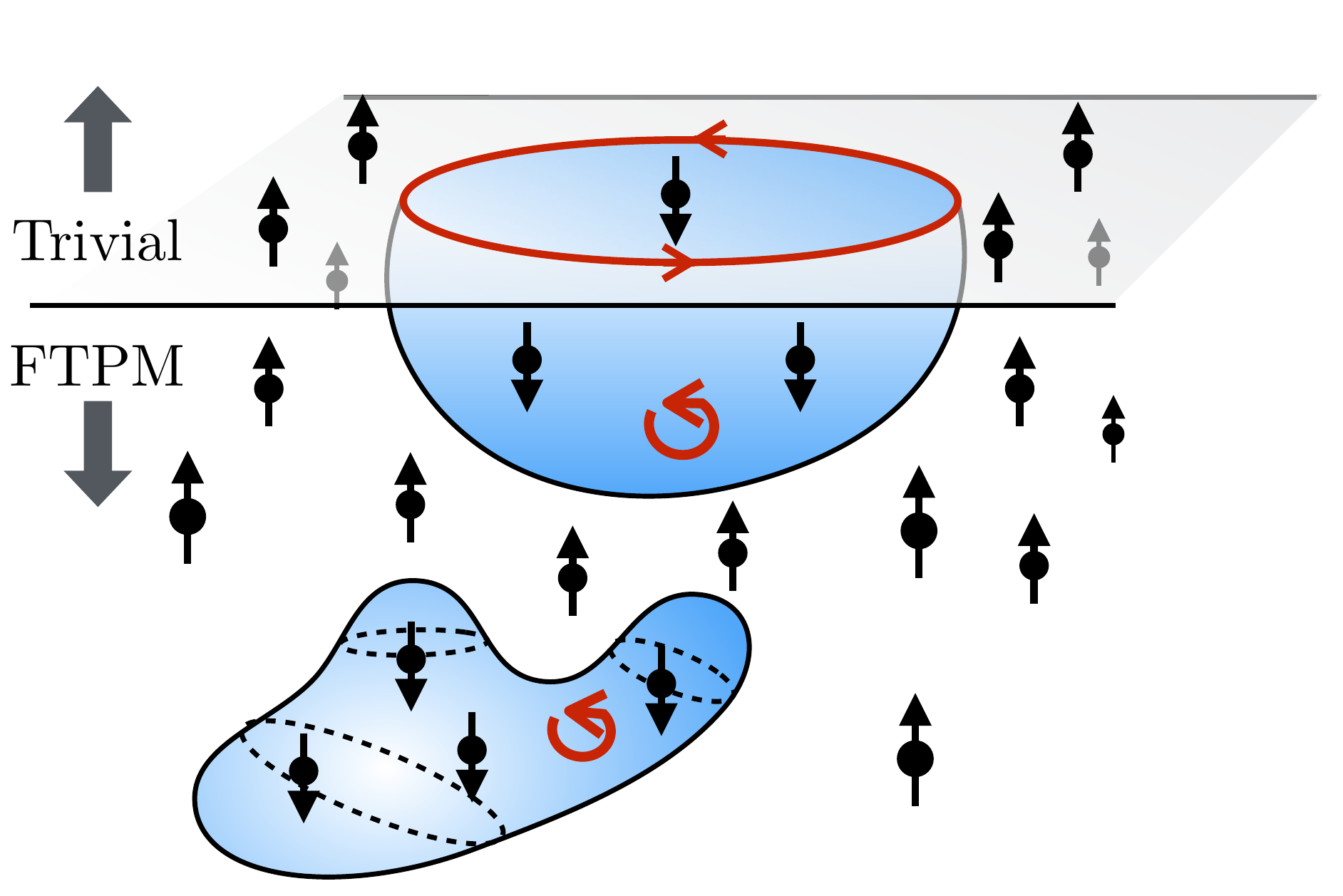}} 
\caption{{\bf Schematic of decorated domain wall construction} Boson degrees of freedom on the domain walls (DWs = blue surfaces) between $\up$ and $\down$ spins (black dots and arrows) are subjected to Chiral Floquet evolution. In the bulk, the DWs form closed surfaces so that the bosons circle around small loops (small red arrow loops). The intersection between a spatial boundary (gray plane) and a DW exposes a long chiral Floquet edge (red circle with arrows).
% In a symmetry preserving surface state, these chiral DWs are proliferated, giving topological protection against surface localization.
\label{fig:model}
 }
\end{center}
\end{figure}

%However, we can modify the above construction, by dividing the boson plaquettes into two groups, labeled I and II, (Fig.~\ref{fig:trs}) based on whether they are oriented along or against an arbitrary fixed vector $\vec{v}$ (which should not point in a direction purely tangential to any boson plaquettes). These two groups are exchanged under TRS. We then apply $U_\text{DDW}$ to group $I$, followed by the time-reversed version of this evolution (which acts on group II), to form a manifestly TRS version of the decorated DW evolution in Eq.~\ref{eq:uddw}:
%\begin{align}
%U(T) =  \(\T U_{\text{DDW},I} \T^{-1}\)^{-1} U_{\text{DDW},I}
%\end{align}
%
%where $U_{\text{DDW},I}$ is obtained by restricting $U_\text{DDW}$ in Eq.~\ref{eq:uddw} to boson plaquettes in type I (which can be accomplished by local spin projection operators). 

%During the first half of the period, the bosons in group $I$ plaquettes evolve according to one cycle of the CF evolution. This results in chiral translation by one unit around the boundary between groups I and II on each DW. During the second half of the period, the $\T$ operators effectively interchange the notion of groups I and II, and together with the inversion, evolve the second group of DW plaquettes according to the same chirality CF evolution (but implemented in a precisely time-reversed fashion compared to the first stage), thereby removing the chiral translation at the seam between groups I and II.

The result is a TRS phase in which the spin DWs are ``decorated" by a CF state with $\nu = \log p$. DWs inside the 3D bulk form closed surfaces, for which the CF evolution is trivial (all bosons traverse short loops and return to their initial position after each period). In contrast, the intersection of a spin DW and the spatial boundary however, exposes the 1D chiral edge state of the CF phase, and produces a non-trivial value of $\eta = p$, and (as we will describe below) topologically protected surface dynamics.

Finally, to convert this idealized zero-correlation length model into a stable MBL phase, we can introduce an extra, disorder step: $U_\text{dis} = e^{-i\sum_i h_i \sigma_i^x + \sum_r\sum_{\alpha = 1}^p \mu_{r,\alpha}|\alpha_r\>\<\alpha_r|}$, where $h_i$ is a random transverse fields for the spins, and $\mu_{r,\alpha}$ gives a random on-site energy to the different states of the $p$-state boson degrees of freedom. This disorder step preserves time-reversal symmetry, and, since $U_\text{DDW}$ is proportional to the identity operator in the bulk, $U_\text{dis}$ preserves the zero-correlation length nature of $U_\text{DDW}$, and also preserves the overall TRS of the evolution.

\header{Surface phases}
With solvable lattice models in hand, we next explore the anomalous surface dynamics of FTPMs. We will see that, while the bulk dynamics of the above model are trivial and localizable, the surface cannot be localized while preserving TRS.  For concreteness, throughout, we will consider an open boundary where the surface terminates on an infinite 2D plane of the $\sigma$-spins.

\begin{figure}[t]
\begin{center}
{\includegraphics[width=0.5 \textwidth]{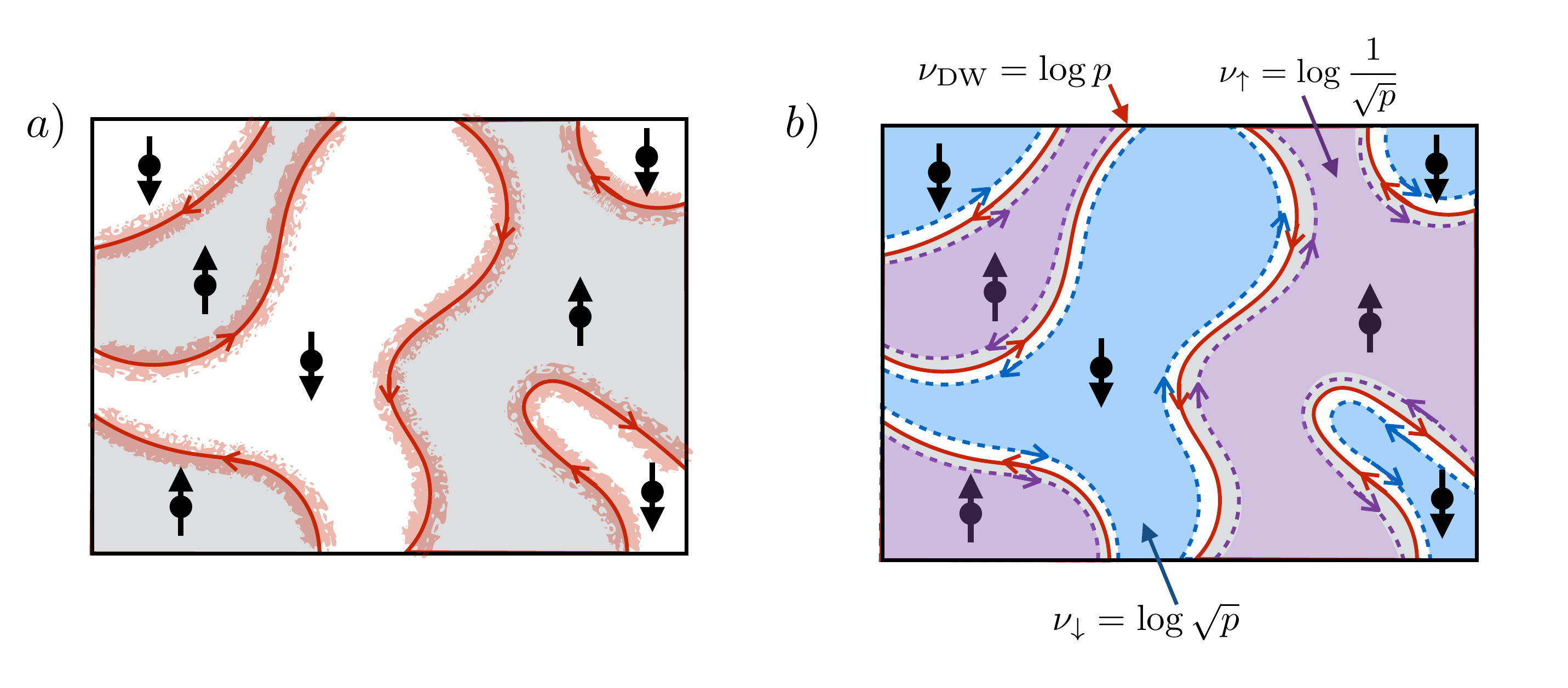}} 
\caption{{\bf Time-reversal symmetric surface phases -- } a) symmetry preserving and thermal due to proliferated chiral edges, b) surface Floquet enriched topological order that is localized with anyonic time-crystal order, due to attaching $2d$ radical chiral Floquet phases to magnetic domains to cancel their chiral motion.
\label{fig:surface}
 }
\end{center}
\end{figure}

\vspace{2pt}\noindent{\it Thermal surface -- }  If we simply extend the bulk evolution all the way to the surface without modification, (which preserves TRS) then the intersection of the spin DWs and the surface carry chiral modes. Then, including the disorder term, $U_\text{dis}$, the surface spins will exhibit a quantum superposition of all chiral DW edges of arbitrarily long lengths (e.g. Fig.~\ref{fig:surface}a), which will necessarily thermalize upon the inclusion of arbitrarily weak perturbations~\cite{po2016chiral}, and cannot be localized, resulting in a delocalized thermal~\footnote{The definition of MBL can be modified to allow for thermal boundaries~\cite{chandran2016many,nandkishore2016general}.} boundary.

\vspace{2pt}\noindent{\it TRS breaking surface -- } Instead, we may localize the surface by breaking TRS, by redefining the projection terms $\Pi$ at in $U_\text{DDW}$ as if the surface layer of spins were perfectly polarized $\up$ in the z-direction (regardless of their actual state). In this case, the CF-coated spin DWs are ``repelled" away from the surface into the bulk and the surface can be fully localized. The time-reversed version of this surface termination would redefine the projectors $\Pi_a$ as if all surface spins were pointing $\down$ in the z-direction. From this construction, one immediately sees that the interface between these two conjugate TRS-breaking boundary configurations has a single chiral Floquet mode corresponding to the edge of a $\nu = \log p$ CF phase, showing that the 3D bulk has $\eta = p$.

\vspace{2pt}\noindent{\it Anomalous surface topological order -- } 
So long as the surface DWs carry CF edge modes with chiral invariant $\nu_\text{DW}$ dictated by the bulk topological properties, then the surface cannot be localized without breaking symmetry. We can attempt to neutralize these chiral channels by ``painting on" 2D CF phases to the TR-breaking surface domains (Fig.~\ref{fig:surface}b). However, in order to preserve TRS, we would have to ``paint" the $\up$ and $\down$ surface domains with TR-conjugate 2D CF phases having chiral invariant: $\nu_\up$ and $-\nu_\down=-\nu_\up$ respectively. Then, all told, the modified surface TR DW would have chiral invariant $\nu_{\text{DW}'} = \nu_\text{DW} + 2\nu_\up$.

 If we could choose the the additional 2D phase to have $\nu_\up = -\frac12\nu_\text{bulk}$, then $\nu_{\text{DW}'}=0$, and we could trivially localize the DW with only a local, TRS modification of the Floquet evolution near the surface. The resulting surface would be both localized and symmetry preserving. However, the resulting surface cannot be topologically trivial. Rather, neutralizing the DW chiral modes in this way would require adding a 2D radical CF phase with $\nu_\up = -\log \sqrt{p}$~\cite{po2017radical,fidkowski2017interacting}. In bosonic systems, such a radical CF invariant is only possible if the drive induces a non-trivial Floquet enriched 2D topological order (FET)~\cite{po2017radical}, exhibiting anyonic excitations that get dynamically exchanged by the Floquet drive. This FET order will persist after the surface DWs have been neutralized and TRS has been restored.  
 
For example, in~\cite{po2017radical} a solvable spin model was constructed that exhibited a radical CF phase with $\nu = \pm \log\sqrt{p}$, which exhibited bulk $\Z_p$ topological order in which the gauge charge and flux excitations were periodically interchanged by the Floquet drive, and whose edges chirally pump non-Abelian parafermionic twist defects with fractional quantum dimension $d=\sqrt{p}$. For the simplest case of $p=2$, the $e\leftrightarrow m$ interchanging $\Z_2$ FET order was shown to always break TRS as indicated by its chiral edge\cite{po2017radical}. However, the above construction shows that this FET order {\it can} occur in a TRS fashion, at the surface of a 3D Floquet TI. This situation is analogous to that of the ordinary equilibrium electronic 3D TI, whose surface can exhibit non-Abelian topologically ordered states similar to the Moore-Read fractional quantum Hall state, which have chiral edge-modes and break TRS when realized in 2D but which can occur without TRS breaking at a 3D TI surface~\cite{wang2013gapped,metlitski2014interaction,chen2014symmetry,bonderson2013time}. We note that our construction of a non-trivial $3d$ surface with $\Z_p$ FET order exchanging $e\leftrightarrow m$ during each period, also shows that this FET order has a dynamical time-reversal anomaly that prevents it from being realized in pure $2d$ settings.

One complication here is that in a generic MBL state with a finite density of $e$ and $m$ excitations, this type of FET order necessarily results in spontaneous breaking of time-translation symmetry\cite{potter2016dynamically,po2017radical} corresponding to a 2T-periodic oscillations between charge and flux anyons. This ``anyonic time-crystal" will arise in the FET phase for for any non-zero density of anyon excitations.

\vspace{2pt}\noindent{\it Time-translation symmetry protection -- } Finally, we note that the topological surface states of FTPMs also rely on the discrete time-translation symmetry associated with the $T$-periodicity of the Floquet drive. For example, their surface states can be trivially localized by $2T$-periodic surface drive, described in Eq.~\ref{eq:Hprime}. We note that there is a formal distinction between spontaneous\cite{else2016floquet}, versus explicit breaking of time-translation symmetry (TTS). Since the invariant $\eta$ is defined in terms of the Floquet evolution operator $U(T)$ itself (rather than its eigenstates), it remains well defined even if the eigenstates of this Floquet operator spontaneously develop motion with an enlarged period~\cite{else2016floquet,khemani2015phase,von2016phaseII,von2016absolute}. However, $\eta$ becomes ill-defined if one introduces perturbations to $U(t)$ that are explicitly $2T$ periodic.

\header{Discussion} 
The infinite family of Floquet topological paramagnets (FTPMs) identified here open an avenue for interacting Floquet topological phases beyond the cohomology framework, with dynamics that cannot be mimicked by any static Hamiltonian system. Extensions of these ideas to fermionic systems and fractionalized phases with topological order is an important task for future work. In Appendix~\ref{app:fermions}, we comment on our current (partial) understanding and open issues for such generalizations.

\renewcommand{\arraystretch}{1.5}
\definecolor{darkred}{rgb}{0.75,0,0}
\begin{table}
\begin{tabular}{C{0.12\textwidth}C{0.08\textwidth}C{0.08\textwidth}C{0.17\textwidth}}
\toprule
Symmetry \& Dimensionality & SPT & X-SPT & F-SPT\\
\midrule
None, $2d$
  & {\color{darkred} $\Z$} & $\Z_1$ (\text{none)} & {\color{blue} $\mathbb{Q}^+\simeq \Z^\infty$}  \\ 
  $\Z_2^T$, $3d$ & $\Z_2 \times${\color{darkred} $\Z_2$}  & $\Z_2$ & $\Z_2\times\blue{\mathbb{Q}^+_2}\simeq \Z_2\times\blue{\Z_2^\infty}$ \\
\bottomrule
%%%%%% 
\end{tabular}
\caption{{\bf Dimensional hierarchy of bosonic SPTs.} Group structure of various types of SPT classifications including equilibrium gapped ground-states (SPT), excited state MBL systems (X-SPT), and periodically driven Floquet systems (F-SPT). The equilibrium SPT include chiral phases with thermal Hall conductance and their $3d$ descendant, the beyond cohomology SPT (red). These cannot be MBL and are absent from the X-SPTs. Instead, for Floquet systems, the chiral phases are replaced by rational CF phases and their $3d$ FTPM descendents SPTs in the second column are the equilibrium gapped ground states, where a time reversal symmetric SPT in $3d$, the beyond cohomology state (blue). Here $\mathbb{Q}^+$ denotes the group of (positive) rationals with multiplication, and $\mathbb{Q}^+_2 = \mathbb{Q}^+/\(\mathbb{Q}^+\)^2\simeq \Z_2^\infty$ is the group of rationals modulo perfect squares. }
 \label{table:spts}
\end{table}

%\begin{table}
%\renewcommand{\arraystretch}{1.5}
%\definecolor{darkred}{rgb}{0.75,0,0}
%\centering
% \begin{tabular}{|c | c | c | c|} 
% \hline
% Symmetry and D & SPT & X-SPT & F-SPT \\ 
% \hline
% No Symmetry, $2d$
%  & {\color{darkred} $\Z$} & $\Z_1$~(\text{none)} & {\color{blue} $\mathbb{Q}^+\simeq \Z^\infty$}  \\ 
% \hline
% $\Z_2^T$, $3d$ & $\Z_2 \times${\color{darkred} $\Z_2$}  & $\Z_2$ & $\Z_2\times\blue{\mathbb{Q}^+_2}\simeq \Z_2\times\blue{\Z_2^\infty}$ \\
% \hline
% \end{tabular}
% \caption{{\bf Dimensional hierarchy of bosonic SPTs.} The SPTs in the second column are the equilibrium gapped ground states, where a time reversal symmetric SPT in $3d$, the beyond cohomology state, is derived from the chiral phases in $2d$ (shown in red). The X-SPTs refer to eigenstate SPT order in excited states of the static problem which can be MBL. Here, the $2d$ chiral states with thermal Hall conductance and $3d$ beyond cohomology states are eliminated, and only the cohomology state remains. Finally the F-SPTs refer to the periodically driven (Floquet) problem. The rational chiral Floquet phases appear instead of the equilibrium chiral ground states and have $3d$ descendants in the presence of time reversal symmetry (shown in blue). Here $\mathbb{Q}^+$ denotes the group of (positive) rationals with multiplication, and $\mathbb{Q}^+_2 = \mathbb{Q}^+/\(\mathbb{Q}^+\)^2\simeq \Z_2^\infty$ is the group of rationals modulo perfect squares. }
% \label{table:SPTs}
%\end{table}

We close by asking how the $3d$ FTPMs fit in within the general set of $3d$ Floquet SPTs of bosons protected by time reversal symmetry and MBL (see Table~\ref{table:spts}). A large class of Floquet SPTs can be understood by applying equilibrium classification techniques (e.g. group cohomology and its generalizations) with an enlarged symmetry group that includes an emergent dynamical discrete time-translation symmetry, $\Z$, in addition to other microscopic symmetries, e.g. for TRS Floquet drives the enlarged group would be $\Z\rtimes\Z_2^T$~\cite{else2016classification,potter2016classification}.
Taking a step back we recall that for the equilibrium case of ground states of gapped bosonic phases, there are two root SPT phases in $3d$, conveniently labeled by their surface topological order, the $eTmT$ and $eFmF$ states~\cite{vishwanath2013physics,wang2013boson}. While the former is captured within group cohomology, the latter is not. However, the $eFmF$ state cannot be MBL, and does not enter the Floquet classification. As we argue in Appendix~\ref{app:generalclassification}, this follows from the fact that the $eFmF$ state is a condensate of $\T$-breaking domain walls decorated by $2d$ chiral E$_8$ states, which exhibit non-zero gravitational anomaly that prevents their localization\cite{kitaev2006anyons,potter2015protection}. Hence, viewing $3d$ TRS Floquet systems as equilibrium systems with an enlarged $\Z\rtimes\Z_2^T$ symmetry group, it would appear that only a single $\Z_2$ invariant (deriving from the equilibrium $eTmT$ state) survives. However, this misses the crucial feature that in $2d$ there are an infinite set of dynamical chiral phases with no equilibrium counterpart, the rational CFs (see Table \ref{table:spts}). These can substitute for the E$_8$ state in decorating the $\T$-breaking domain walls -- leading to the infinite family of 3d FTPMs discussed in this paper.

\vspace{0.1in}
\noindent{\it Acknowledgements -- } We thank H.-C. Po, D. Else and D.S. Freed for helpful conversations. ACP is supported by NSF DMR-1653007. LF is supported by NSF DMR-1519579, Sloan FG-2015- 65244. AV acknowledges support from a Simons Investigator Award and AFOSR MURI grant FA9550-14-1-0035. This research was supported in part by the Kavli Institute of Theoretical Physics and the National Science Foundation under Grant No. NSF PHY11-25915.

\appendix

\section{Details of lattice model construction \label{app:model}}
In this Appendix, we provided details of the construction of the lattice models realizing the decorated domain wall Floquet evolution described in the main text. We first construct a convenient lattice implementation of a $2d$ CF phase, which forms the basis for the domain wall decoration in the $3d$ lattice models of FTPMs.

\subsection{$2d$ Chiral Floquet Model}
To build chiral Floquet (CF) evolution $H_{\text{CF},s,a}(t)$ utilized in the main text, we need to choose a particular implementation of the $2d$ chiral Floquet unitary evolution of the $p$-state bosons.
%, and which can be applied on arbitrary planar domains. 
Previously constructed CF models based on applying a time-dependent sequence of boson SWAP operations\cite{rudner2013anomalous,po2016chiral} have inconvenient properties for arranging onto arbitrary $2d$ planar domains, and we will find it convenient to design an alternative (though topologically equivalent) implementation, whose edge acts as a uniform chiral edge-translation by one site, and which can be easily applied to a $2d$ domain of arbitrary geometry.

A key building block in this construction is an operator: $C_{a,s}$ that cyclically permutes the p-state bosons around a square plaquette, $a$, in either a right ($s=+1$) or left ($s=-1$) handed sense. For example, for a four site plaquette, $a$, with sites labeled $_A^D \square_B^C$:
\begin{align}
C_{a,+} = \sum_{j_{A\dots D}=1}^p |j_D,j_A,j_B,j_C\>\<j_A,j_B,j_C,j_D| \equiv e^{-iH_aT}
\end{align}
The cyclic permutations, $C_{a,s}$ can always be generated by a local Hamiltonian, $H_{a}$, i.e. $C_{a,s} = e^{-isH_{a}T}$, acting only on the spins in plaquette $a$.
%,  such that $\T H_{a,s} \T^\dagger = -H_{a,s}$, i.e. time reversal switches the orientation of the symmetry: , and $\T C_{a,s}\T^\dagger = C_{a,s}$ (note, that this indicates that $C_{a,s}$ is time-reversal \emph{odd}, since a time-reversal symmetric unitary evolution, $U_\text{TRS}$, would satisfy $\T U_\text{TRS} \T^{-1} = U_\text{TRS}^\dagger$.) To see this, note that We can explicitly write $H_{a,s}$, by identifying the eigenstates of $C_{a,+}$: $C_{a,+}|n\> = \lambda_n|n\>$, and then writing $H_{a}$  as a sum over projectors: $H_{a} = \frac{1}{T}\sum_n i\log \lambda_n |n\>\<n|$. We note that, due to the time-reversal odd nature of $C_{a,+}$, the eigenvalue associated with $|n\>$ gets complex conjugated by time-reversal, i.e. $C_{a,+}\T|n\> = \lambda^*_n|n\>$. Then, by choosing an appropriate branch of the $\log$, we can ensure that $H_{a}$ is time-reversal odd: $\T H_{a}\T{-1} = -H_{a}$. Since $\lambda_n^4 = 1$, then $\lambda_n$ will always be of the form $e^{2\pi i N/4}$ for integer $N$, and we can choose a branch such that $i\log \lambda_n \in \{0,\pm \pi/2, \pm \pi\}$, i.e. $i\log \lambda_n^* = -i\log \lambda_n$. 

\begin{figure}
\begin{center}
\includegraphics[width=0.3 \textwidth]{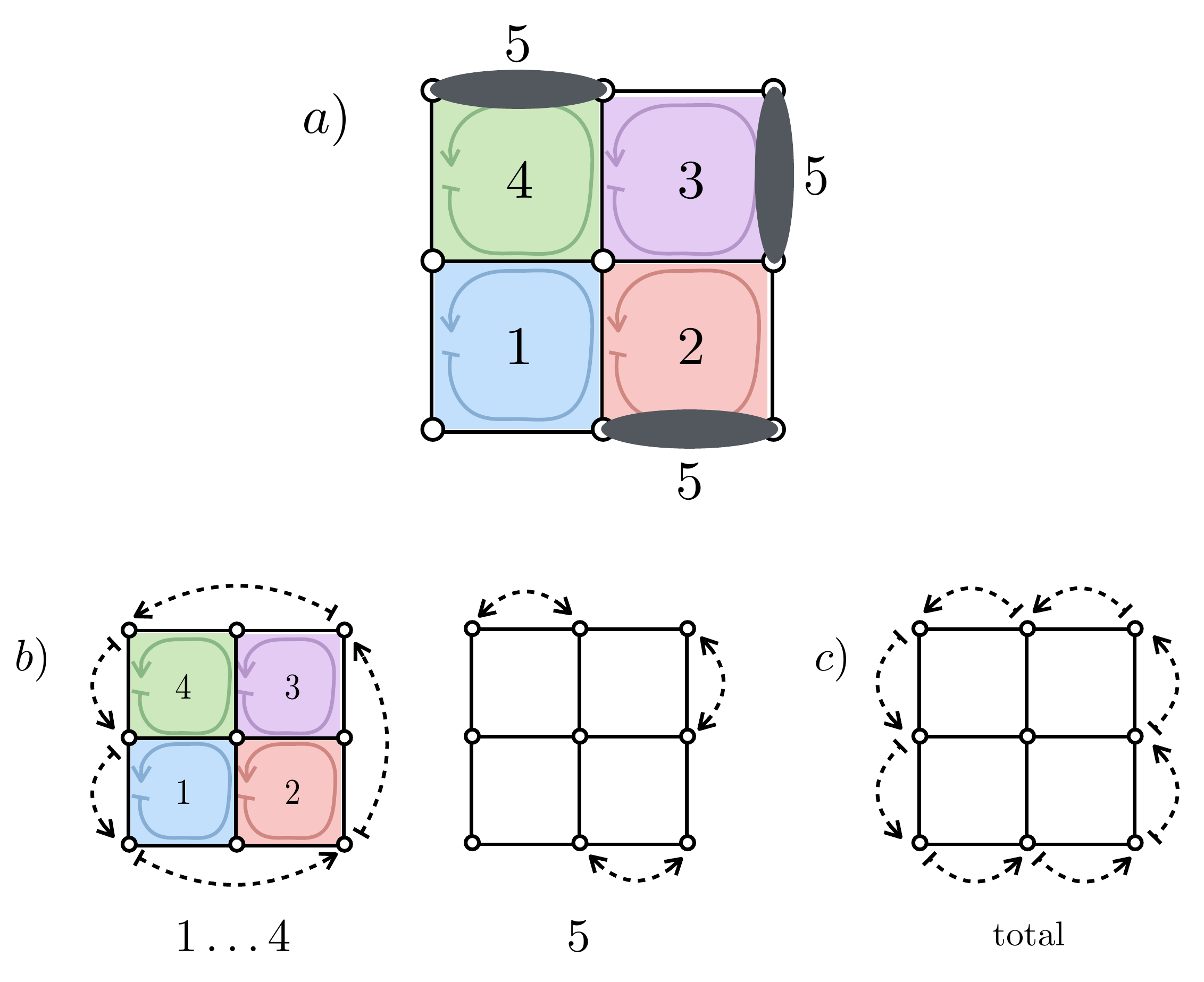}
\\ \vspace{.2in}
\includegraphics[width=0.2 \textwidth]{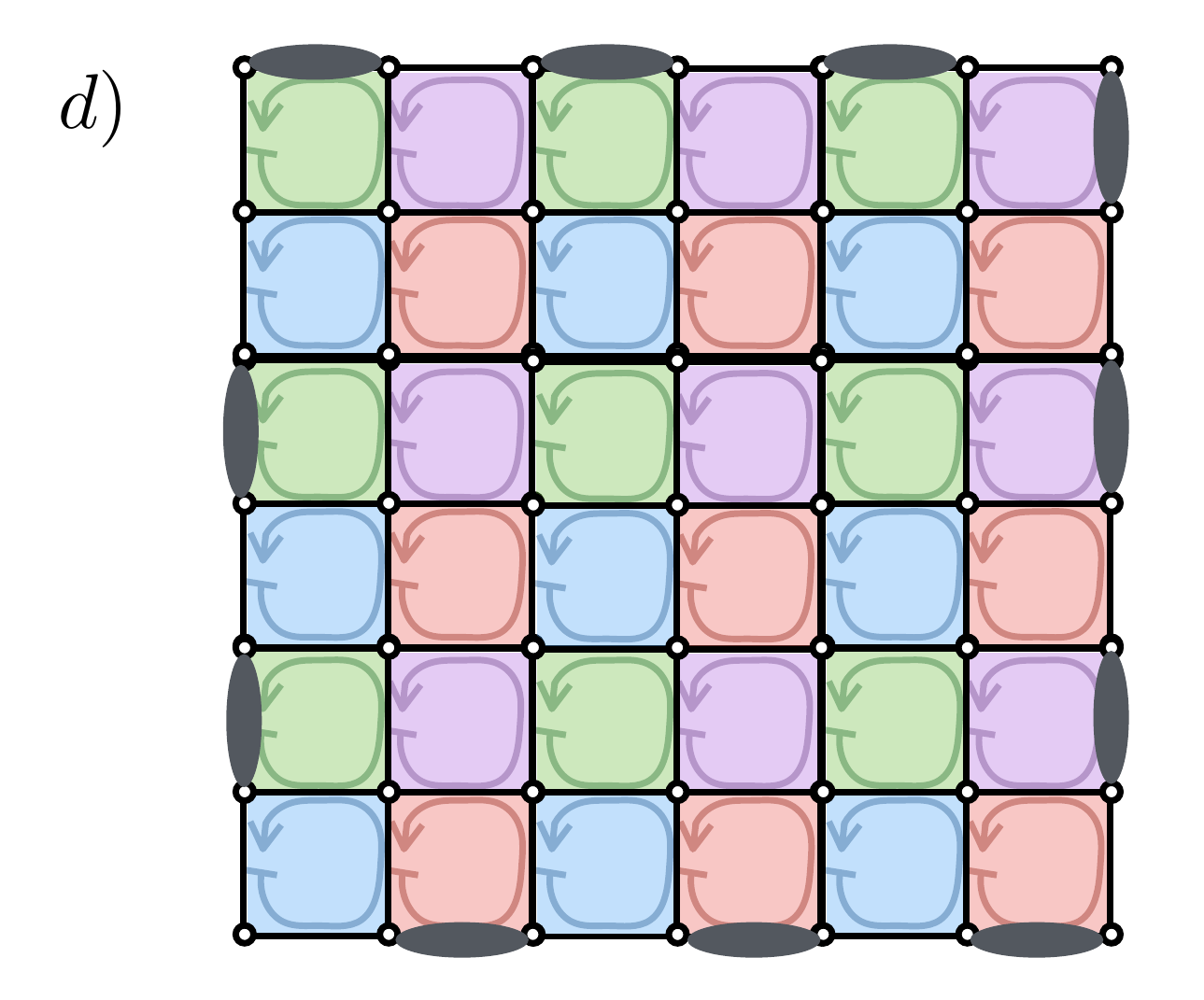}
\caption{{\bf 2d chiral floquet model} a) Schematic of the 5-step $2d$ lattice model with CF evolution for a 9-site boson plaquette. Steps $1\dots 4$ consist of applying chiral permutations (colored circling arrows) around the 4-site boson to the plaquettes of type 1 (blue), 2 (red), 3 (green), and 4 (purple) in sequence. Steps $1\dots 4$, are equivalent to a uniform chiral edge translation up to a finite depth local unitary transformation (panel b). This difference is undone in step 5 by appropriately chosen boson SWAP operations (gray ovals), such that the total evolution results in uniform edge translation by one site (panel c). Panel e) illustrates the 5-step evolution for a larger $7\times 7$-site square. 
\label{fig:2dcf}
 }
\end{center}
\end{figure}

To implement the CF evolution on the spin DWs, we can then label all of the plaquettes of the boson lattice by a number between $1$ and $4$ (see Fig.~\ref{fig:2dcf}), and sequentially apply $C_{a,s}$ on plaquettes of type $1$, $2$, $3$, and then $4$. The result of this four-step sequence is shown in Fig.~\ref{fig:2dcf}b, for a $3\times 3$ square. There is no motion for the site at the center of the square. The states of the edge sites are moved either $0$, $1$, or $2$ sites along the edge in a chiral fashion (dashed arrows in Fig.~\ref{fig:2dcf}b), such that one boson state is transferred across each point along the edge. This evolution differs from an ideal chiral edge translation only by a finite-depth local unitary evolution. We can remove this superficial difference by applying a 5th stage of the evolution, in which the boson states are swapped between neighboring sites shown with a dark gray oval in Fig.~\ref{fig:2dcf}a. 

\begin{figure}
\begin{center}
\includegraphics[width=0.4 \textwidth]{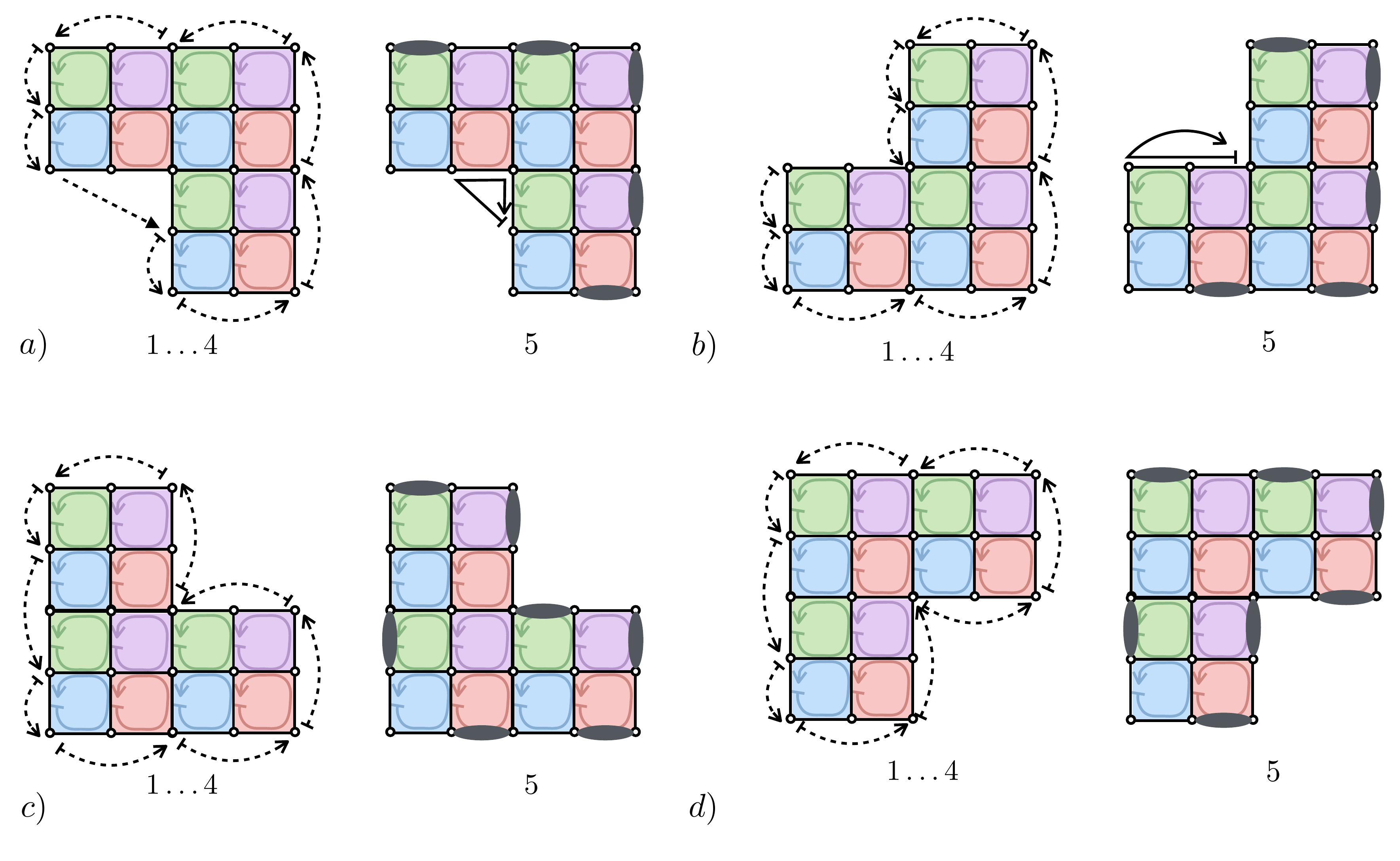}
\includegraphics[width=0.35 \textwidth]{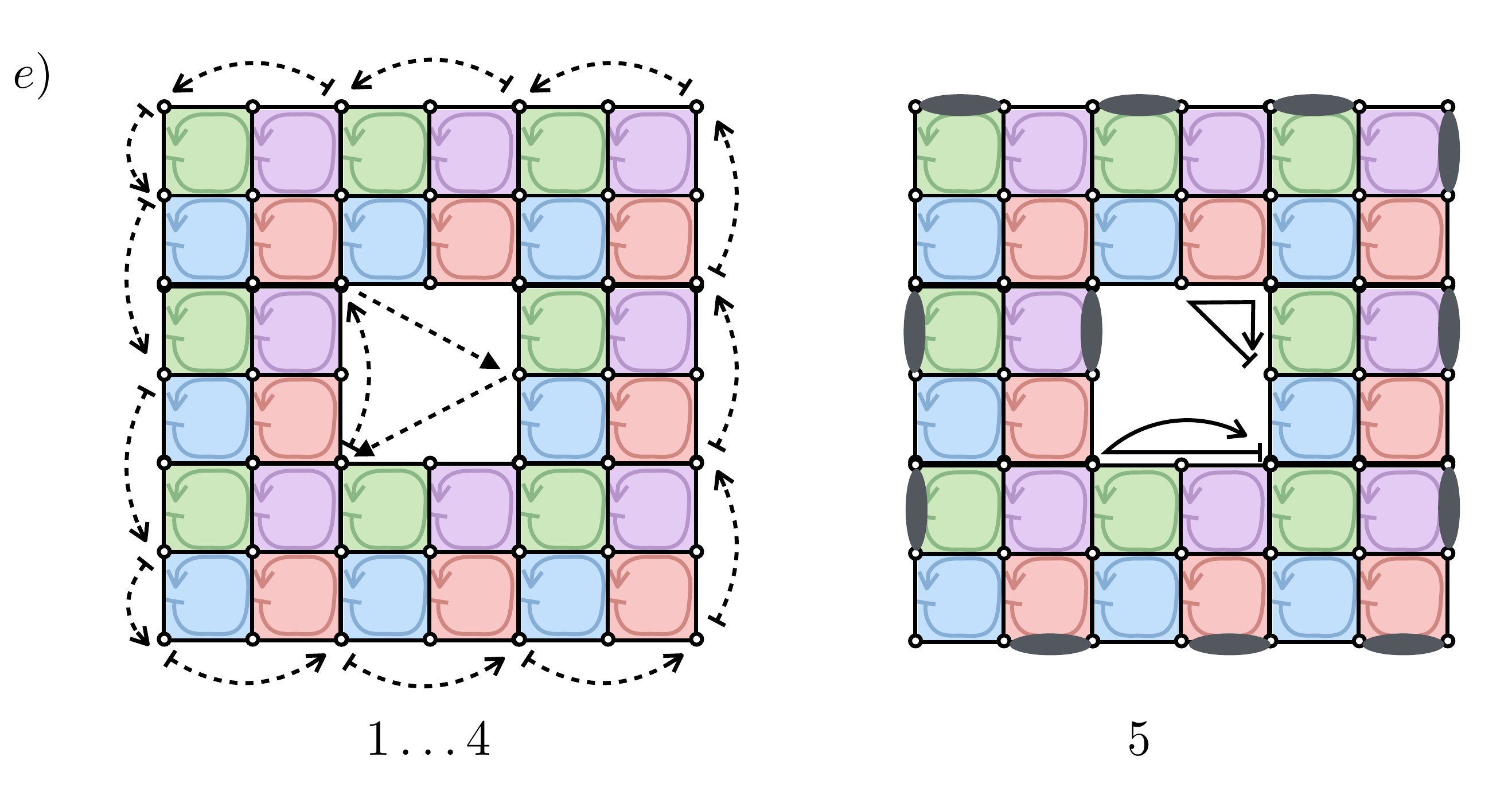}
\caption{{\bf 2d chiral Floquet model on non-square domains} 
\label{fig:2dcfexamples} The difference in the edge motion (dashed arrows) from stages $1\dots 4$ and the ideal uniform chiral edge translation can be removed by a local unitary transformation (step 5), that depends on the local geometry of the $2d$ domain, and involves either 2-state bosonic SWAP operations (gray ovals) or 3-state chiral SWAP operations (cyclic solid arrows). (a)-(e) show illustrative examples for various domain shapes made from edge-sharing tilings of the minimal ``unit cell" shown in Fig.~\ref{fig:2dcf}a), that consists of $4$ square boson plaquettes.
 }
\end{center}
\end{figure}

The resulting 5-step evolution implements an idealized chiral edge translation unitary, in which the bosons at the boundary are shifted by precisely one site in the right-handed direction. This idealized CF evolution can be implemented on arbitrary $2d$ domains made from arbitrary edge sharing configurations of this minimal $3\times 3$ unit. The precise implementation of step $5$ depends on the local geometry of the $2d$ domain (illustrative examples are given in Fig.~\ref{fig:2dcfexamples}).

\subsection{$3d$ Decorated Domain Wall Model}
With the $2d$ CF implementation in hand, we can begin to assemble the $3d$ decorated domain wall model of a FTPM. The lattice model is formed from two types of degrees of freedom:
\begin{enumerate}
\item Spins-1/2, $\vec{\sigma}_i$ that transform under TRS as $\T\vec{\sigma}_i\T^{-1} = \sigma^x_i\vec{\sigma}_i^*\sigma^x_i$, and
\item p-state bosons, with an onsite Hilbert space spanned by a basis: $\{|1\>,|2\>\dots|p\>\}$ that transforms trivially under TRS.
\end{enumerate}

\begin{figure}
\begin{center}
a) \includegraphics[width=0.4 \textwidth]{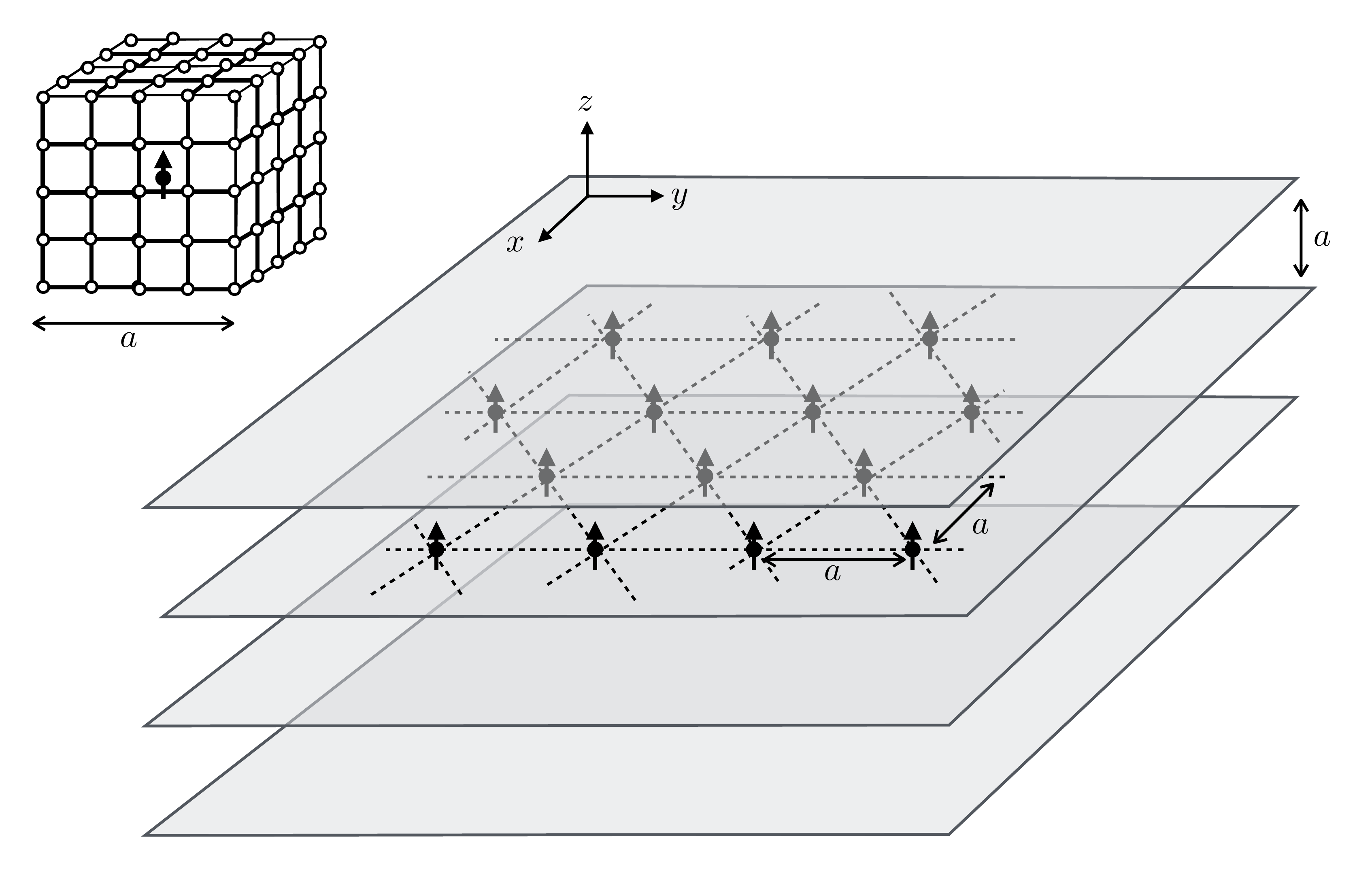}
%\\
b) \includegraphics[width = 0.3\textwidth]{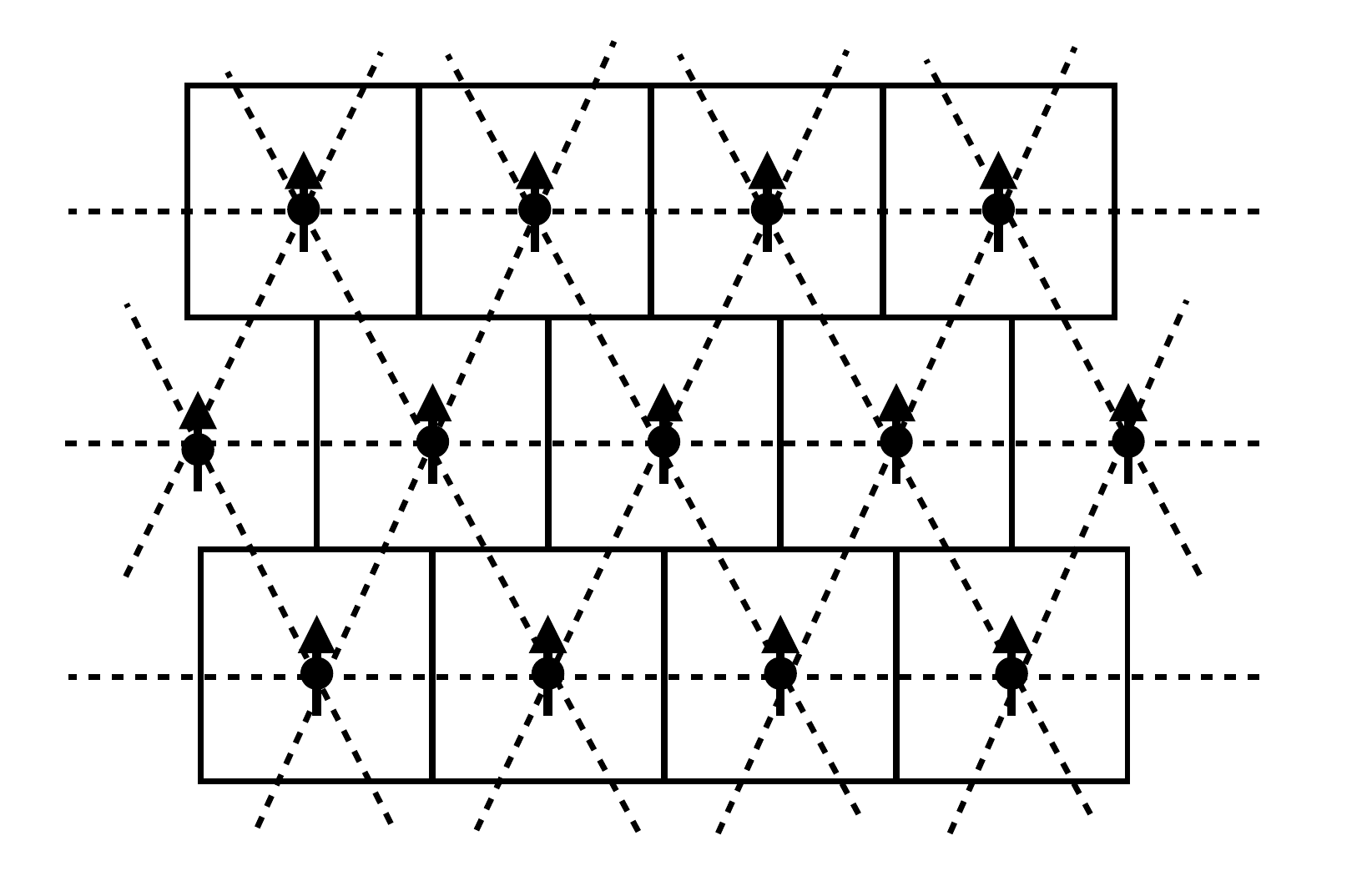}
\caption{{\bf Schematic of 3d decorated domain wall model -- }  (a) The $\sigma$-spins (black arrows) form a layered triangular lattice (layers depicted as gray sheets). Each spin is surrounded by a cube of boson sites (inset), with each face of the cube containing a grid of $p$-state boson sites (open circles). (b) Top view of one of the layers, with the boson cubes (black squares, see inset of panel a for detailed structure of each cube) forming a brick lattice surrounding the triangular lattice of spins.
\label{fig:ddwmodel}
 }
\end{center}
\end{figure}

We arrange the spin-1/2, $\sigma$ degrees of freedom on a layered triangular lattice with each layer being a vertically shifted copy of the one below it (Fig.~\ref{fig:ddwmodel}). We take each $\sigma$-spin to be surrounded by a cube with a $5\times 5$ grid of $p$-state boson sites on each face. In each layer, the boson cubes form a brick lattice around the spins, which has the advantage that the spin domain walls projected onto the boson cube faces will always contain an integer number of the elementary $4$-square-plaquettes used in the $2d$ CF implementation (Fig.~\ref{fig:2dcf}). This arrangement also avoids issues with point-like intersections between domain walls.

Our strategy will be to apply the $2d$ CF evolutions described in the previous section, to the boson plaquettes sitting on spin DWs. We define an orientation of the DWs point from down to up spins, and will evolve with the CF phase of right-handed chirality (with respect to the DW orientation). We further label the minimal 4-square boson plaquettes based on the direction of their normal vector: $\pm x$, $\pm y$, or $\pm z$ (with $\pm$ sign given by the DW orientation). 

\begin{figure}[t]
\begin{center}
\includegraphics[width=0.5 \textwidth]{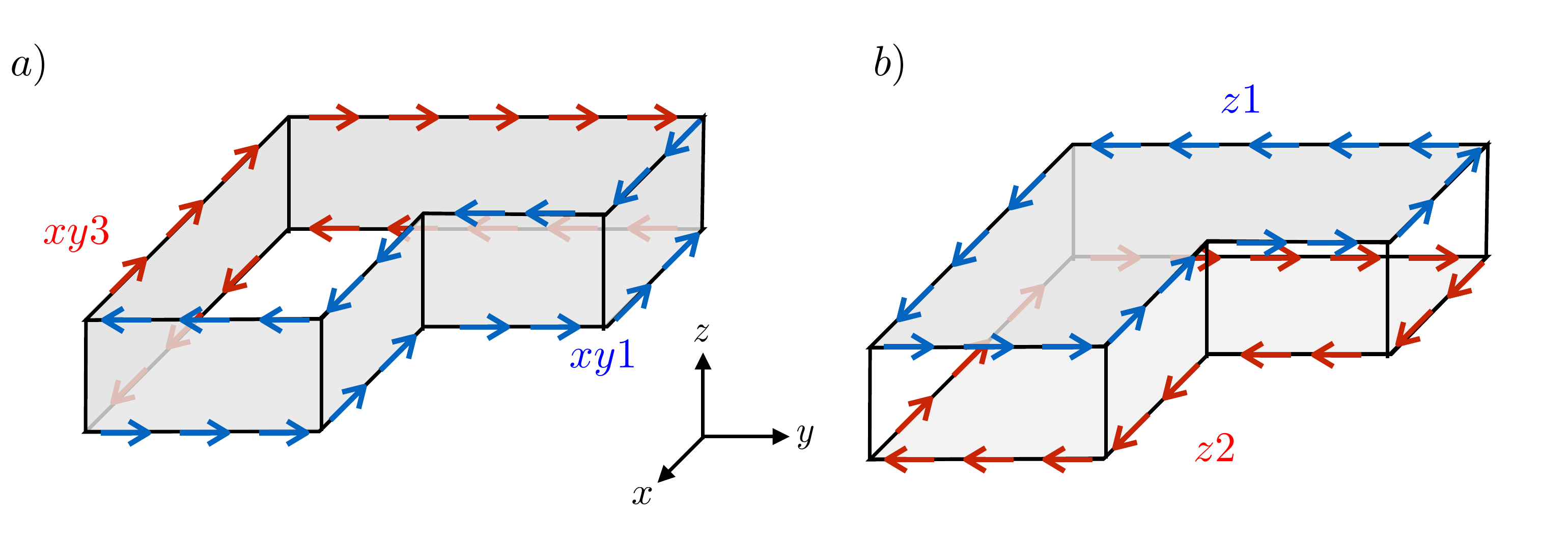}\\
\caption{{\bf Two stage DDW evolution -- }  The solid shape represents a spin domain wall (DW) with spin down inside and spin up outside.The decorated domain wall (DDW) evolution proceeds in two stages:  First (a), the xy-oriented boson plaquettes residing on spin domain walls (gray shaded plaquettes in panel a) undergo a CF evolution. To ensure TRS, this CF evolution is implemented in three steps as described in the main text. The effect of the steps xy1 and xy3 are indicated by blue and red arrows respectively. Second (b), the z oriented boson plaquettes on the spin domain walls (gray shaded plaquettes in panel b) undergo a CF evolution, again in two steps z1 and z2 (whose effects are indicated by red and blue arrows in panel b).
\label{fig:xyz}
 }
\end{center}
\end{figure}

Complications arise in ``folding" the $2d$ CF evolution onto a closed $3d$ surface. Namely, it is impossible to evolve all of the $x$, $y$, and $z$ plaquettes simultaneously according to the 5-step CF evolution, without acting with $C_{a,+}$ on multiple overlapping plaquettes at the same time. Since $C_{a/b}$ commute only for disjoint plaquettes $a$ and $b$, this would spoil the desired zero-correlation length property of the model (i.e. render it not exactly solvable). To avoid this problem, we divide the CF evolution into two stages, evolving the $x$ and $y$ plaquettes in the first stage, and then the $z$ plaquettes in the second stage (Fig.~\ref{fig:xyz}). This division enables us to apply a $C_a$'s to a disjoint set of boson plaquettes at every step.

\begin{figure}[t]
\begin{center}
\includegraphics[width=0.4 \textwidth]{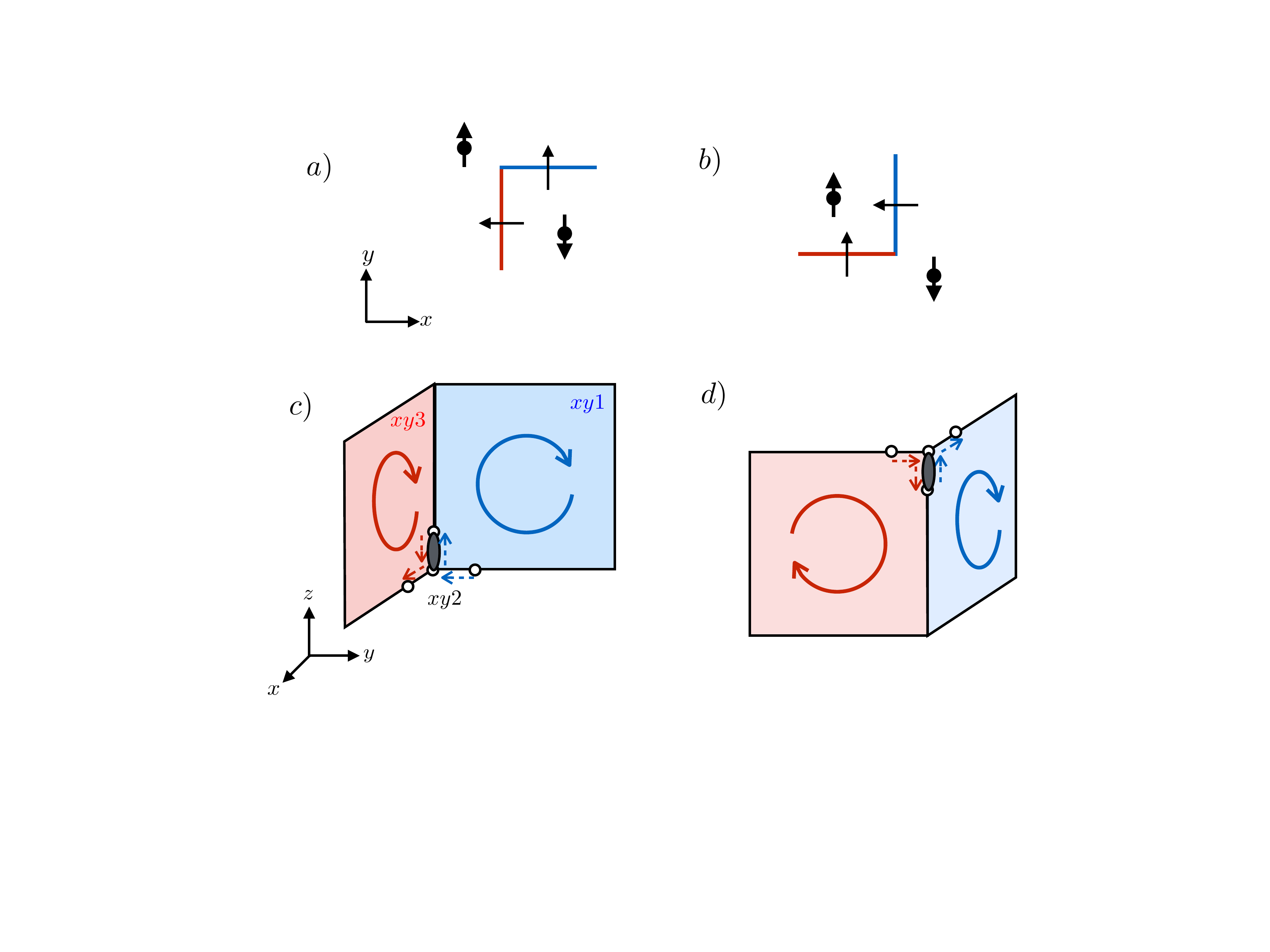}
\caption{{\bf Evolution near corners of xy-plaquettes -- } Top view (a,b) and perspective view (c,d) of two corners of the intersection between $\pm \hat{x}$ and $\mp \hat{y}$ plaquettes, whose CF evolution needs to be patched together by an extra SWAP operation (gray oval). Domain walls are oriented (black arrows) from spin down to up (balls and arrows). In the first step, xy1, the $+\hat{x}$ and $+\hat{y}$ oriented plaquettes (blue) are evolved with the CF evolution with right-handed chirality indicated by the circular arrow, then the two site SWAP operations are applied in step: xy2, and finally the time-reverse of the first step is applied to the remaining $-\hat{x}$ and $-\hat{y}$ oriented plaquettes (red). Motion (dashed arrows) is indicated for the CF steps only for the boson sites (open circles) near the problematic corners.
\label{fig:corner}
 }
\end{center}
\end{figure}

In order to make the overall Floquet evolution TRS, we will need to further sub-divide the first stage into three steps:
\begin{enumerate} % start: xy step details
\item[\bf xy1:] Evolve the $+\hat{x}$ and $+\hat{y}$ facing plaquettes with the CF evolution. Denote this unitary evolution as $U_{xy1}$.
\item[\bf xy2:] Apply an appropriate set of SWAP operations to such that step xy1, and the next step, xy3, result in ideal chiral translations at the boundaries between xy and z surfaces (see Fig.~\ref{fig:corner}) 
\item[\bf xy3:] Evolve with $U_{xy3}= \(\T^{-1} U_{xy1}\T\)^{-1}$
\end{enumerate}

Step xy3 is effectively the same as applying the CF evolution to the $-\hat{x}$ and $-\hat{y}$ oriented plaquettes. To see this, note that the time-reversal operators flip the spin-projectors in $U_{xy1}$ so that $\T^{-1} U_{xy1}\T$ implements a left-handed CF evolution as if the spin domain orientation were reversed. Hence, this will act on the negative $xy$-oriented plaquettes that were left out of step xy1. Lastly, the overall inversion in $U_{xy3} =   \(\T^{-1} U_{xy1}\T\)^{-1}$ switches the CF evolution back to the original right-handed one (though still acting on the $-x$ and $-y$ oriented plaquettes). Together with an appropriate choice of the SWAP operations in step xy2 (see Fig.~\ref{fig:corner}), xy3 undoes the chiral motion at the boundary of the $+\hat{x}$ and $+\hat{y}$ plaquettes, leaving only a chiral motion around the z-plaquettes (which subsequently be undone in the second, $z$, stage of the evolution).

The virtue of dividing the xy-evolution into these steps is that it ensures that the overall evolution for the xy-stage: 
\begin{align}
U_{xy} = U_{xy3}U_{xy2}U_{xy1}
\end{align}
is manifestly time-reversal symmetric. Specifically, since $\T U_{xy1} \T^{-1} = \T \(\T^{-1} U_{xy1}\T\)^{-1} \T^{-1} = U_{xy1}^{-1}$, and since the SWAP operations used in $U_{xy2}$ are manifestly TRS ($\T U_{xy2} \T^{-1} = U_{xy2}^\dagger$), we verify: $\T U_{xy}\T^{-1} = U_{xy}^\dagger$.

To complete the construction, we need to remove the remaining chiral motion at the edge between the xy-facing and z-facing DW plaquettes. This is done by applying right-handed CF evolutions to the z-plaquettes. Again, to ensure TRS, it is convenient to divide this z-stage into two steps:
\begin{enumerate} % start: z-step details
\item[\bf z1:] Evolve the $+\hat{z}$ facing plaquettes.
\item [\bf z2:] Evolve with $U_{z2} = \(\T^{-1} U_{z1}\T\)^{-1}$, which does the CF evolution of the same orientation to the $-\hat{z}$ facing plaquettes in a way that is manifestly the time-reverse of step z1.
\end{enumerate} % end: z-step details
Finally, in order to combine the $xy$ and $z$ stages together in a way that is overall TRS, we should ``sandwich" the z-steps around the the xy steps as:
\begin{align}
U(T) = U_{z1} U_{xy} U_{z2}
\end{align}

This unitary evolution implements the decorated domain wall picture of the FTPM phase described in the main text, while preserving the overall TRS. We note, in passing, that $U(T)$ is unitarily equivalent to applying all the steps in sequence as shown in Fig.~\ref{fig:xyz}: $U_{z2}U_{z1} U_{xy} = U_{z2} U(T) U_{z2}^\dagger$. Changing between these two orderings simply amounts to a shift in our definition of the period, and a corresponding shift to the  ``center of inversion" for the time-reversal operation. We can readily verify that the construction of $U(T)$ results in a TRS evolution:
\begin{align}
\T U(T) \T^{-1} &= \T U_{z1}\T^{-1} U_{xy}^\dagger \T U_{z2} \T^{-1} = U_{z2}^{\dagger} U_{xy}^\dagger U_{z1}^\dagger
\nonumber\\  &= U(T)^\dagger
\end{align}

\begin{figure}
\begin{center}
\includegraphics[width=0.5 \textwidth]{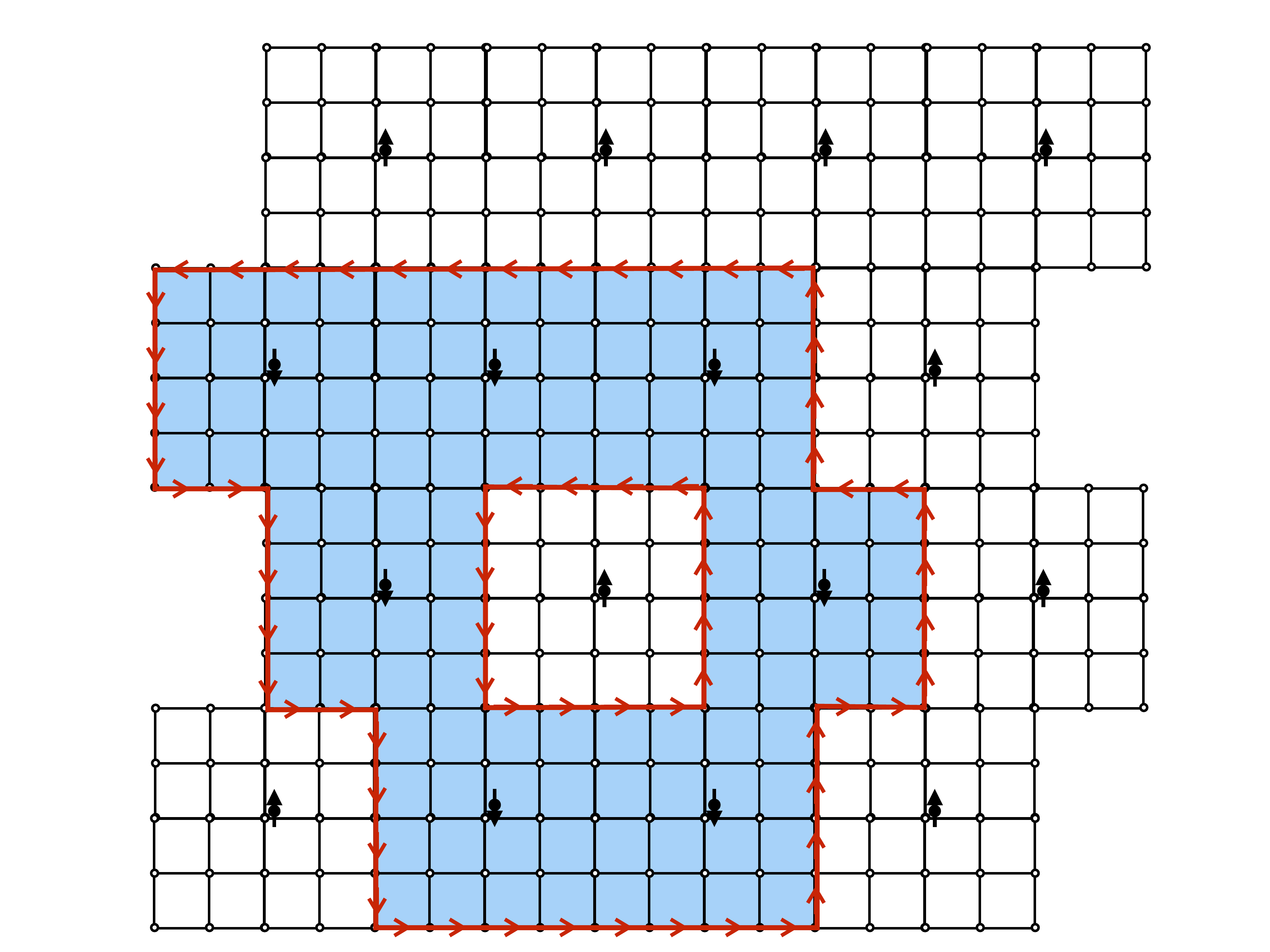}
\caption{{\bf Surface chiral domains -- } The intersection (blue area) of a DW between $\down$ and $\up$ $\sigma$-spins and a spatial boundary exhibits chiral translation of the $p$-state bosons. Spins are represented by black arrows, and are recessed into the page by a half-lattice spacing. The boson sites (open circles) exhibit chiral translation (red arrows) around the spin domains.
\label{fig:surfacedw}
 }
\end{center}
\end{figure}

Crucially, each of these steps can be implemented by a local Hamiltonian involving projectors onto spin configurations multiplied by the local boson terms corresponding to the terms in the appropriate $2d$ CF implementation described in the previous section. The evolution is easiest to picture when projected onto a $2d$ plane. Fig.~\ref{fig:surfacedw} shows the result of the $z$ plaquette evolution for a fixed arrangement of spins. One can readily verify that, overall, every bulk site on the DW returns to itself after one period of this evolution (even at the corners and edges of various $3d$ domain shapes, Fig.~\ref{fig:corner}), so that the Floquet evolution is equivalent to the identity in the bulk. However, the intersection of a DW with a spatial boundary exposes a chiral edge (Fig.~\ref{fig:surfacedw}), as required for the FTPM phase.

\section{Possible fermionic analogs \label{app:fermions}}
In this section we investigate the possible connections of the ideas in the main text to fermionic SPT phases protected by TRS, whose surface states are characterized by time-reversal breaking domain walls that exhibit the chiral edge dynamics of a $2d$ fermionic CF phase\cite{fidkowski2017interacting}. Our current understanding of such fermionic generalizations is incomplete at present, and this appendix aims to assemble our current partial knowledge, lay out possible scenarios, and highlight open issues.

\subsection{Decorated domain walls with spinless fermions}
A seemingly simple extension of the models described in the main text is to replace the $p$-state boson sites with complex (spinless) fermions described by creation operators $f^\dagger_r$ on site $r$. Since the on-site Hilbert space of a fermion has the same number of states as a $p=2$ boson, this procedure would naively also produce a phase with $\eta = 2$. In the presence of microscopic fermions, though, this $\eta=2$ fermion phase seemingly requires charge conservation symmetry in addition to time-reversal symmetry for stability, since in superconducting systems, a purely $2d$ Majorana CF phase with $\nu = \sqrt{2}$ is possible\cite{po2016chiral,po2017radical,fidkowski2017interacting}.

This construction appears to yield a Floquet topological insulator (FTI) protected by $U(1)$ charge conservation and spinless time-reversal symmetry, $\T$, with $\T^2=1$ (class AI). We will refer to this phase as a ``spinless fermion FTI".

 In the absence of interactions, FTIs are described by a non-interacting Floquet band structure, whose possible topological features have exhaustively classified\cite{roy2016periodic}. For this class, there are no non-trivial Floquet-band TI phases, i.e. the spinless fermion FTI does not exist without interactions. Hence, we are left with three possibilities:
\begin{enumerate}
\item The spinless fermion FTI is topologically equivalent to a purely bosonic $\eta=2$ phase,
\item the spinless fermion FTI is inequivalent to any bosonic phase, and instead constitutes an intrinsically interacting fermion Floquet SPT phase, or
\item there is a subtle (i.e. currently unknown) way in which the spinless fermion FTI surface states are not topologically protected.
\end{enumerate}
While we do not currently have a definitive understanding of which option is correct, let us weigh some circumstantial evidence regarding each of these possible scenarios.

Two observations speak in favor of scenario 1 (spinless fermion FTI = bosonic FTPM). First, we can employ a gedanken experiment that is frequently useful in equilibrium SPTs of inserting a $\pi$-flux (vison) into the surface state. In equilibrium, a non-trivial fermionic SPT will react non-trivially to such a $\pi$-flux (e.g. the flux will acquire a symmetry protected degeneracy or fractional symmetry charge)  -- otherwise one could proliferate such $\pi$-flux excitations and gap out the fermions at the surface, showing that the topological properties arise from purely bosonic degrees of freedom. The spinless fermion FTI phase, on the other hand, does not have a topological response to a $\pi$-flux, suggesting a topological equivalence to a purely bosonic system. A second piece of circumstantial evidence for scenario 1, is that in the absence of $U(1)$-number conservation symmetry, the $\nu = \log 2$ CF phase of bosons and fermions are topologically equivalent\cite{po2016chiral,fidkowski2017interacting}.

However, there are two possible reasons to doubt these arguments (supporting scenario 2, spinless fermion FTI $\neq$ bosonic FTPM). 
First, the application of the $\pi$-flux proliferation to ``gap" out the fermion degrees of freedom is subtle in the context of highly excited states of a Floquet MBL system where energy is not conserved, and the topological properties come from highly excited dynamics, potentially involving excitations with arbitrary quasi-energy. Second, the demonstration of equivalence between boson and fermion CF phases with $\nu = \log 2$ in Ref.~\cite{po2016chiral}, hinged on the absence of $U(1)$ charge conservation, to show that the boson and fermion on-site Hilbert spaces can be made equivalent by tacking on auxiliary degrees of freedom with trivial dynamics. One can readily convince themselves that this trick cannot be done in a charge conserving way, so long as one has a finite on-site Hilbert space. Namely, in a fermion system, all bosonic degrees of freedom have even charge, whereas all fermionic degrees of freedom have odd charge. Hence, the maximal charge state in a fermionic site can never be equivalent to that of a bosonic site. This raises the more subtle possibility, that the fermion and boson phases may only be equivalent in a system with an infinite on-site Hilbert space (e.g. a quantum rotor model), though such an unbounded on-site Hilbert space may be problematic for MBL. 

Lastly, while we presently see no concrete issue with the fermionic decorated domain wall (DDW) model, there is potential cause to worry that there is some hidden obstruction that we have yet to identify (scenario 3). For example, in equilibrium, one could try to create a DDW model of an ordinary electronic TI, by decorating TRS-breaking magnetic domains with integer quantum Hall states of spinless fermions with $\sigma^{xy} = \frac{e^2}{h}$. This would seemingly result in a model in which surface magnetic domains have a single chiral mode equivalent to the quantum Hall edge -- the hallmark of an electronic TI with electromagnetic theta angle $\theta_e=\pi$. However, in that context, it is known that the surface state is not protected, and that only spinful (Kramers doublet, $\T^2=-1$) electrons can form a stable TI phase. By analogy, it is conceivable that the ``spinless FTI" is not a stable topological phase, but rather, its surface state is not SPT protected. Instead, a non trivial SPT order requires spin-1/2 fermions to form. Such a spinful fermion FTI (class AII) does exist in the absence of interactions, and is characterized by a non-trivial Floquet band invariant\cite{roy2016periodic}. This spinful fermion FTI cannot be many-body localized without breaking time-reversal symmetry due to local Kramers degeneracies\cite{potter2016symmetry}, however, it may occur as a long-lived pre-thermal phenomena. 

At the present, we are unable to definitively decide among these three scenarios, and raise this task as a challenge for future work.

\subsection{Spinful fermion Floquet topological insulator}
In this section, we briefly outline the topological properties of a (non-interacting or prethermal) Floquet topological insulator of spin-1/2 electrons protected by charge conservation and spinful time-reversal symmetry (class AII). We will call this phase the ``Floquet band TI". This phase is classified by a non-trivial Floquet band invariant\cite{roy2016periodic}. In general, Floquet band structures are classified by two copies of the equilibrium band invariants. The first copy can be intuitively viewed as an equilibrium phase that is realized in a Floquet context, and has the usual equilibrium topological surface states at quasi-energy $0$. Similarly, the extra Floquet phases can be viewed as a second set of equilbrium invariants for topological surface states at quasi-energy $\pi$. 

This leads to an intuitive picture of the Floquet band TI surface states, as consisting of one Dirac cone $0$ quasi-energy and another at $\pi$ quasi-energy (strictly speaking, since there is no particle-hole symmetry, it is the quasi-energy difference between the two surface Dirac cones that is fixed at $\pi$). Namely, as was shown in Ref.~\cite{yao2017discrete} (see also \cite{von2016absolute}), for any phase in which time-evolution by two periods can be implemented by an ordinary local, symmetric, and time-independent Hamiltonian $U(2T) = e^{2iH_\text{eff}T}$ one can formally define an emergent dynamical $\Z_2$ symmetry generator: $g = U(T)e^{iH_\text{eff}T}$ where $U(2T)$. The two surface Dirac cones have opposite symmetry ``charge" under this emergent dynamical symmetry.

We can further understand the properties of this phase, via a gedanken experiment in which we insert a minimal flux magnetic monopole into the Floquet band TI. The familiar topological index theorems for Dirac cones guarantee that each surface Dirac cone contributes a charge-1/2 fermionic bound state to the monopole, described by annihilation operators $\psi_{0,\pi}$ respectively. Crucially, the quasienergy difference between these modes is precisely: $\delta \e = \pi$. The charge neutral monopole has two possible configurations, where one of the $\psi_{0,\pi}$ is occupied and the other empty. These two configurations differ in quasi-energy by $\pi$, or equivalently, the two states of the neutral monopole are degenerate with respect  to time evolution by two periods, $U(2T)$. This dynamical ``degeneracy" is protected by emergent dynamical $\Z_2$ symmetry that descends from the time-translation symmetry of the drive. 

We close by remarking that this monopole gedanken experiment distinguishes the spinful Floquet band TI, from the putative spinless  fermion FTI explained in the previous section. The latter does not respond in any non-trivial way to $U(1)$ fluxes, and cannot be topologically equivalent to the Floquet band TI. 

\subsection{Fractionalized generalizations}
In the main text and previous sections, we have focused on short-range entangled bulk phases without anyon excitations, we may also consider consider ``fractional" analogs of these FTPM phases which can be accessed via a related decorated domain wall construction in which TRS breaking domains are decorated with radical CF phases, which would result in intrinsic 3D bulk topological order. The surface of a putative fractional DTI would then have an effective fractional value of the 3D TRS topological invariant $\eta \in \prod_i \(\sqrt{p_i}\)^{n_i~\text{mod}~2}$, and surface states with effective chiral index that is a quartic root of a rational number, $r$, $\nu_\text{surface} = \pm\log \sqrt[4] r$. While such phases should be stable as metastable pre-thermal ``ground-states", there is a potential complication for realizing an MBL state in disordered versions of these systems. Namely, the bulk 3D topological order would exhibit string-like gauge-flux excitations, which, in the idealized zero-correlation length limit, would result in an exact degeneracy growing exponentially with the number of intersections between the string excitations and fluctuating spin-DWs. Upon moving away from the fine-tuned integrable limit, in the related case of 2D radical CF phases, these degeneracies were lifted either by a spontaneous breaking of time-translation symmetry, or a breakdown of MBL. Whether simply breaking time-translation symmetry in the above construction is sufficient to produce a stable MBL phase remains an open question for future study.

\section{General classification of bosonic Floquet topological paramagnets \label{app:generalclassification}}
Recall that the static equilibrium classification of 3d bosonic phases with time reversal symmetry is $\Z_2 \times \Z_2$, with one $\Z_2$ generated by the in-cohomology ($eTmT$) SPT state~\cite{chen2013symmetry}  and the other by the beyond-cohomology ($eFmF$) SPT state~\cite{vishwanath2013physics}.  In this appendix we will argue that the in-cohomology state can be realized by a many-body localizable (MBL) Hamiltonian, whereas the beyond-cohomology one cannot.  A general proof that all in-cohomology states are MBL was given in~\cite{potter2015protection}. Here, we present a related, complementary argument that also allows us to argue that the beyond-cohomology state cannot be localized. Thus, the proposed full classification of bosonic Floquet topological paramagnets will include the in-cohomology SPT state, together with the infinite family of models constructed in this paper: i.e. the new infinite family replaces the beyond-cohomology state in the Floquet classification

They key property of any in-cohomology SPTs in spatial dimension $d \geq 1$ is the fact that its ground state can be disentangled into a symmetric product state by a finite depth unitary $V$ that commutes with all symmetry generators:
\begin{align}
|\Psi_{g.s.}\rangle = V |\Psi_{prod}\rangle.
\end{align}
Here $|\Psi_{g.s.}\rangle$ is the SPT ground state, $V$ is a finite depth circuit of local unitary operators, and
\begin{align}
|\Psi_{prod}\rangle = \otimes_j |\psi_j\rangle
\end{align}
is a tensor product state over the sites $j$ of the system of symmetric states $|\psi_j\rangle$.  Indeed, the disentangling circuit $V$ can be constructed directly for the zero correlation length models introduced in \cite{chen2013symmetry}, and its existence is a universal property of the SPT phase. This ground-state construction was generalized to all excited states of an MBL system in \cite{potter2015protection}. Note that only the entire circuit $V$ is symmetric -- the individual unit-depth unitary steps making up $V$ will not, by themselves, be symmetric for a non-trivial SPT phase. 

If the symmetry group $G$ is onsite and Abelian, then each site Hilbert space decomposes as a sum of orthogonal one dimensional representations $\alpha$.  Letting $P_j^\alpha$ denote the projector onto the $\alpha$ representation at site $j$, we see that the ground state $|\Psi_{g.s.}\rangle$ is the unique state annihilated by $\{V P_j^0 V^\dag\}$, where $0$ denotes the symmetric representation generated by $|\psi_j\rangle$.  Thus the Hamiltonian

\begin{align}
H_{\text{MBL}} = \sum_{j,\alpha} J_{j,\alpha} V P_j^{\alpha} V^\dag
\end{align}
with suitably chosen (random or quasi-periodic) couplings $J_{j,\alpha}$ is a Hamiltonian with a full set of local conserved quantities believed to be in or close to an MBL phase, realizing $|\Psi_{g.s.}\rangle$ as an eigenstate.

Conversely, if an SPT ground state can be realized as the eigenstate of a symmetric Hamiltonian with a full set of symmetric local conserved quantities, then we conjecture that such a symmetric disentangling circuit $V$ must exist.  Indeed, in this case one can define $V$ via the unitary transformation mapping projectors onto on-site degrees of freedom to projectors onto the conserved quantities (``l-bits") of the MBL system, which is a locality preserving, and quasi-local transformation. Indeed, \cite{gross2012index} showed that in one dimension such locality-preserving unitary operators are always finite depth quantum circuits up to a generalized translation.  Assuming some version of this result holds in higher dimensions, and the generalized translation can be argued to act trivially (i.e. take product states to product states), this would imply the existence of a symmetric finite depth circuit disentangling the ground state.  

Although we cannot prove this at this time, we conjecture that this is true in the case of 3 spatial dimensions and time reversal symmetry. In fact, we can almost take the existence of such a $V$ as the definition of symmetry preserving MBL. Certainly any counterexamples would have a very different structure than known MBL phases, and would likely require modifying several common definitions of MBL. In particular, the in-cohomology, $eTmT$ state has a TRS disentangling circuit $V$ and can be many-body localized in this fashion.  On the other hand, we claim that for the beyond cohomology state, no such circuit $V$ exists, which by the argument above strongly suggests that it cannot be MBL.

We will now argue by contradiction, that no such circuit exists for the beyond cohomology, $eFmF$, state. Suppose that a symmetric circuit, $V$ did exist, which could disentangle the bulk of the $eFmF$ state in the absence of boundaries. Then, consider a system with a boundary (e.g. a solid rectangular block with a surface), and as the surface state, take the time reversal symmetric $eFmF$ state \cite{vishwanath2013physics,burnell2013exactly}.  This is a gapped surface topological order with an anomalous realization of time reversal symmetry: namely, any truly 2d realization of the $eFmF$ state necessarily has a chiral central charge equal to $4$ modulo $8$.

In this open geometry, we can define a truncation of the hypothetical $V$ to the bulk, which disentangles the bulk but not the boundary (generically it is not possible to disentangle the boundary of a nontrivial SPT with a finite depth unitary).
%version of the putative $V$ by rest Then consider a large system with a boundary, e.g. a solid block with a surface, and define Since $V$ is only defined for systems without boundary, in this case we let $V$ denote a truncation of the disentangling circuit to the bulk of the system: i.e. $V$ will not disentangle the boundary.  
The putative $V$ would be TRS in the bulk. Namely, writing the time-reversal operator as $\T = U_{\T}K$ with $U_{\T}$ being a product of on-site unitary operators and $K$ being complex conjugation (in some basis), $V$ and $U_{\T}^\dag V^*  U_{\T}$ have the same action on operators localized in the bulk of the system.  Note that time reversal property of the disentangling circuit $V$ is not the same as that of the time evolution operator, which requires an extra inverse.  Thus we see that the operator $V^{-1} U^\dag V^* U$ acts only on the spins localized near the surface, i.e. is a surface operator.

The key point now is that $V$ would break time reversal symmetry near the surface and maps an SPT eigenstate to a product state in the bulk tensored with a surface state $|\Psi_s\rangle$. This surface state $|\Psi_s\rangle$ is now a truly 2d realization of the $eFmF$ state, and hence has a nonzero chiral central charge of $c$ equal to $4$ modulo $8$.  On the other hand, if we had disentangled the bulk of the original (TRS) state with the time-reversed partner $U^\dag V^* U$ of $V$, we would have obtained the time-reversed $eFmF$ state, with chiral central charge $-c$.  Also, these two time-reversed incarnations of the $eFmF$ surface state are mapped into each other by the 2d locality preserving surface operator $V^{-1} U^\dag V^* U$.

This leads to a contradiction, as follows.  Let us denote by $eFmF_+$ and $eFmF_-$ the surface states with chiral central charge $c$ and $-c$ respectively, and let $W=V^{-1} U^\dag V^* U$ be the 2d locality preserving operator that maps one to the other.  Now stack each of $eFmF_+$ and $eFmF_-$ with another copy of $eFmF_-$.  Augmenting $W$ by the identity on this second copy of $eFmF_-$, we obtain an operator $\tilde{W}$ that maps $eFmF_+ \times eFmF_-$ to $eFmF_- \times eFmF_-$.  However, $eFmF_+ \times eFmF_-$ is simply the quantum double of $eFmF$, and necessarily has a parent Hamiltonian equal to a sum of local commuting projectors.  Conjugating these local commuting projectors by the locality-preserving operator $\tilde{W}$ we would obtain a local commuting projector Hamiltonian for $eFmF_- \times eFmF_-$, which is impossible because $eFmF_- \times eFmF_-$ has nonzero chiral central charge\cite{kitaev2006anyons}.  This is the desired contradiction.

%The defining property of the $eFmF$ phase is that $|\Psi_s\>$ is necessarily equivalent to an odd (in particular non-zero) number of $E_8$ phases.  Then the operator $V^{-1} U^\dag V^* U$, which acts only near the surface and is hence a $2d$ locality preserving unitary would map the surface state, $|\Psi_s\rangle$, to a state with the opposite thermal Hall conductance.  In other words, we would have constructed a 2d locality preserving unitary operator that takes a state with one value of thermal Hall conductance to a state with a different value of thermal Hall conductance. However, it is well known that changing the Hall conductance is not possible with a finite depth unitary circuit\cite{kitaev2006anyons}, providing the desired contradiction.

%However, then by stacking with an appropriate number of $E_8$ states we would be able to construct a locality preserving surface unitary operator that takes a state with zero thermal Hall conductance to one with non-zero thermal Hall conductance.  But a state with zero thermal Hall conductance and no symmetry requirements can always be disentangled by a finite depth circuit, and hence has a parent commuting projector Hamiltonian.  Conjugating each projector by our locality preserving surface unitary, we would then obtain a commuting projector Hamiltonian for a state with non-zero thermal Hall conductance, which is impossible \cite{kitaev2006anyons}.

\bibliography{FloqSPTbib}
\end{document}